\documentclass{elsarticle}
\usepackage{lineno,hyperref}

\usepackage{chemformula} 

\usepackage{graphicx} 
\usepackage{subcaption} 
\usepackage[separate-uncertainty = true, multi-part-units = single, list-units = single, range-units = single]{siunitx} 
\usepackage{comment}

\DeclareSIUnit\Molar{\textsc{m}}
\DeclareSIUnit\rpm{rpm}
\DeclareSIUnit\ppm{ppm}

\DeclareSIUnit\kbt{k_BT}

\usepackage[noabbrev,nameinlink]{cleveref}
\usepackage{multirow}
\usepackage{makecell}
\usepackage{colortbl}
\usepackage{amsmath}
\usepackage{amssymb}
\usepackage{xr}
\usepackage{siunitx}

\usepackage{setspace}
\usepackage{blindtext}
\usepackage{geometry}

\usepackage{ragged2e}

\usepackage{lineno}
\justifying

\begin{document}

\begin{frontmatter}

\title{Exploiting anisotropic particle shape to electrostatically assemble colloidal molecules with high yield and purity}

\author[1]{Yogesh Shelke}
\author[2]{Susana Mar\'in-Aguilar}
\author[2]{Fabrizio Camerin}
\author[2]{Marjolein Dijkstra}
\author[1]{Daniela J. Kraft\corref{cor1}}

\ead{kraft@physics.leidenuniv.nl}
\cortext[cor1]{Corresponding author}

\address[1]{Soft Matter Physics, Huygens-Kamerlingh Onnes Laboratory, Leiden University, PO Box 9504, 2300 RA Leiden, The Netherlands}
\address[2]{Soft Condensed Matter, Debye Institute for Nanomaterials Science, Utrecht University, Princetonplein 1, 3584 CC Utrecht, The Netherlands}

\begin{abstract}

{\textbf{Hypothesis:} Colloidal molecules with anisotropic shapes and interactions are powerful model systems for deciphering the behavior of real molecules and building units for creating materials with designed properties. While many strategies for their assembly have been developed, they typically yield a broad distribution or are limited to a specific type. We hypothesize that the shape and relative sizes of colloidal particles can be exploited to efficiently direct their assembly into colloidal molecules of desired valence. 

\textbf{Experiments:}  We exploit electrostatic self-assembly of negatively charged spheres made from either polystyrene or silica onto positively charged hematite cubes. We thoroughly analyze the role of the shape and size ratio of particles on the cluster size and yield of colloidal molecules.  

\textbf{Findings:} Using a combination of experiments and simulations, we demonstrate that cubic particle shape is crucial to generate high yields of distinct colloidal molecules over a wide variety of size ratios. We find that electrostatic repulsion between the satellite spheres is important to leverage the templating effect of the cubes, leading the spheres to preferentially assemble on the facets rather than the edges and corners of the cube. Furthermore, we reveal that our protocol is not affected by the specific choice of the material of the colloidal particles. Finally, we show that the permanent magnetic dipole moment of the hematite cubes can be utilized to separate colloidal molecules from non-assembled satellite particles. Our simple and effective strategy might be extended to other templating particle shapes, thereby greatly expanding the library of colloidal molecules that can be achieved with high yield and purity.}
\end{abstract}


\end{frontmatter}


\section{Introduction}

Colloidal molecules are clusters of colloidal particles with features that resemble those of real molecules. Their anisotropy in terms of shape and interactions is strikingly similar to that of their molecular analogues. This makes them powerful model systems for unraveling the behavior of real molecules,~\cite{duguet2011design,poon2004colloids} for designing building units to create structures with desired properties,~\cite{he2020colloidal,aryana2019superstructures} and for serving as a versatile basis for creating active particles with a more complex motion behavior than has been observed for simple spheres.~\cite{soto2014self,ni2017hybrid,alvarez2021reconfigurable,lowen2018active}

Various methods for the production of colloidal molecules with different shapes, sizes, and functionalities have been developed in the past years, employing for example synthetic strategies such as nucleation and growth,~\cite{sun2015controllable} seeded emulsion polymerization,~\cite{tian2022particle} physical vapor deposition,~\cite{schamel2013chiral} emulsion-based clustering,~\cite{cho2005colloidal,ku2015soft}, colloidal fusion,~\cite{gong2017patchy} and self-assembly approaches.~\cite{kurka2020self,kraft2012surface} While bottom-up synthetic routes have the advantage of producing large quantities of uniform colloidal molecules, the obtained shape and interaction complexity is often limited. In contrast, self-assembly strategies typically provide access to more complex particle shapes and interactions. Assembly can be driven by a manifold of approaches ranging from the hybridization of surface-bound~\cite{soto2002controlled} or surface-mobile DNA linkers,~\cite{chakraborty2022self}  hydrophobic~\cite{chen2011supracolloidal} and depletion interactions,~\cite{sacanna2010lock} to the assembly driven by opposite charges.~\cite{maansson2019preparation,demirors2015opposite,mihut2017assembling, liu2021assembly}
On the other hand, they also require strategies to limit the assembly to finite-sized clusters with a predetermined valence,~\cite{bianchi2017limiting,li2020colloidal,ni2016programmable,wang2012colloids} a mechanism to achieve a uniform arrangement of the constituent particles in the clusters, and separation of the resultant colloidal molecules by their size or type.

A straightforward self-assembly approach for producing colloidal molecules from binary mixtures of colloidal spheres is hetero-aggregation through electrostatic interactions.~\cite{soto2002controlled, schade2013tetrahedral, mihut2017assembling, demirors2015opposite, liu2021assembly} In this procedure, two particle species with opposite surface charge are selected and mixed such that they aggregate in solution. This aggregation process can be considered irreversible, meaning that when the particles are linked they fluctuate at most near their contact position. To limit the growth of the resulting structures, core particles of type A are typically used in the  presence of an excess of satellite particles of type B. Upon assembly, the excess type B colloids saturate the surface of the less frequent type A particles, thereby halting the electrostatically driven aggregation and yielding finite-sized clusters. These so-called colloidal molecules can  be characterized by the coordination number $N$ of B-type particles that are attached on a single A-type particle as AB$_{N}$ complexes. 

For spheres, the number of attached particles $N$, and thus the size and shape of the resulting colloidal molecule, depends largely on their size ratio of the two types of spheres.~\cite{miracle_influence_2003,maansson2019preparation,hu2020particle} While the geometry in principle sets the maximum cluster size,\cite{miracle_influence_2003} in experiments typically a wider range of cluster types is obtained.
The reason for this is that the assembly is a random sequential aggregation process and immobility of the outer particles after adhesion to the core particle prevents optimization of space and hence reaching the maximum number of outer particles bound to the core.~\cite{tagliazucchi_kinetically_2014, schade2013tetrahedral} As a result, the yield for a specific valence is typically low, except for AB$_4$ clusters assembled at a critical size ratio $\alpha_c=2.41$ and AB$_2$ clusters for  $\alpha\geq 6.46$.~\cite{schade2013tetrahedral} 
In contrast, if polyhedral metal-organic framework particles were used as the center of the cluster, it was found that their geometry in principle sets the cluster size.~\cite{liu2021assembly} For sufficiently high number ratios, the polyhedral shape allows for adsorption of one sphere per facet if the adsorbing spheres are neither too large nor too small. However,  how the size ratio is connected to the successful assembly of highly coordinated clusters and thus whether this strategy is robust to polydispersity was not investigated. In addition, other electrostatic effects such as repulsion between the outer particles might influence the outcome.  

For hetero-coagulation of oppositely charged spheres, it is known that electrostatic interactions can play a role in determining the cluster size. Depending on the screening length of the solvent and the charges of both particles, clusters may have a higher or lower coordination number.~\cite{demirors2015opposite} Self-assembly again occurs via random sequential aggregation with particle immobility after adhesion, implying that no optimization of the sphere configuration is possible after adhesion and hence that the yield of a given type of colloidal molecule is limited as well. Although strategies using density gradient centrifugation~\cite{manoharan2003dense} and fluorescence-activated cell sorting instruments~\cite{mage2019shape} are available, selective separation of a single colloidal molecule type is often time-consuming and challenging. 

Using a combination of experiments and simulations, we here demonstrate that a non-spherical particle shape, a cube, can significantly enhance the formation of colloidal molecules of a specific type for a wide range of size ratios. Specifically, using electrostatically driven hetero-coagulation of negatively charged spheres (B) and positively charged cubic core particles (A), we show that the cubic shape favors the self-assembly of colloidal molecules with AB$_2$ and AB$_6$ across a wide range of size ratios for different materials of spheres.  
We furthermore demonstrate that AB$_4$-type colloidal molecules can be obtained by confining the assembly to quasi-2D by sedimentation, using satellite particles of a high density material. The size ratio of the particles together with the non-spherical particle shape limits the maximum number of bound particles and can be used to produce large quantities of either of these favored species over a wide range. Using numeric simulations, we find that electrostatic repulsion between the bound spheres plays a crucial role in the cluster formation, by favoring aggregation on the facet of the cubic core over adsorption to the  edges and corners of the cube. 
Clusters assembled from hematite cubes and polystyrene spheres can be further stabilized by fusing the polystyrene satellites using Tetrahydrofuran (THF) as plasticizer.~\cite{hueckel2018mix} Finally, we demonstrate that the permanent magnetic dipole of the employed hematite cubes can be used to straightforwardly separate colloidal molecules from the excess spheres through a magnetic field.

\section{Materials and Methods}

\subsection{Experimental Section}
\subsubsection{Materials}
Polystyrene spheres with 2.00±0.05 $\mu$m diameter and 5 wt/v$\%$ were purchased from Sigma-Aldrich. Polystyrene spheres with 4.55±0.05 $\mu$m diameter and 10 wt/v$\%$ were purchased from  Microparticles GmbH. Polystyrene spheres of diameter 1.00±0.05 $\mu$m were synthesized by surfactant free emulsion polymerization.~\cite{wel2017surfactant} 
Silica spheres of diameter 0.97±0.05 $\mu$m,  2.06±0.05 $\mu$m and 4.62±0.05 $\mu$m size with 5 wt/v$\%$  were purchased from Microparticles GmbH. Sodium hydroxide (NaOH, 99.5$\%$ from Acros Organics), tetrahydrofuran (THF, 99.9$\%$ from Honeywell Research Chemicals), and Iron(III) Chloride Hexahydrate (FeCl$_3 \cdot$ 6H$_2$O) (ACS, 97.0-102.0$\%$ from Alfa-Aeser) were used as received. Milli-Q (Millipore Gradient A10) water with resistivity 18.2 M$\Omega \cdot$cm was used for all experiments.

\subsubsection{Synthesis of Hematite Cubes}
The hematite cubes were synthesized using the procedure described by Sugimoto et al.~\cite{sugimoto1992preparation} In a typical synthesis that yielded 0.83±0.03 $\mu$m hematite cubes, 100 ml 2M FeCl$_3 \cdot $ 6H$_2$O was prepared in a 500 ml pyrex bottle. Next, 100 ml of a 5M NaOH solution was added in 20 seconds while stirring. Then, the mixture was kept stirring for another 10 minutes and subsequently placed and kept undisturbed in a preheated oven at 100 $^\circ$C for 8 days. After that, the resulting cubes were washed several times using centrifugation and re-dispersion in Mill-Q water. Finally, the particles were first washed with ethanol and then with water to remove unreacted chemicals by repeated centrifugation at 5000 rpm for 10 min (Beckman Coulter, Avanti J-15R) and re-dispersion by sonication (Elma, Elmasonic P) for 15 min at 20 $^\circ$C temperature and 37 MHz frequency. A similar procedure and concentration were used for synthesizing 1.08±0.04 $\mu$m hematite cubes except that the 100 ml of 5M NaOH solution was added in 40 seconds while stirring. The cube's size was determined by measuring the side to side length of at least 100 particles from scanning electron microscopy (SEM) using ImageJ. For imaging, a 50 $\mu$l suspension of hematite cubes of concentration of 0.05 wt/v$\%$  was spread and dried on a glass substrate at room temperature. After sputter coating the substrate with a 5nm platinum layer (Cressington 208HR sputter coater), the sample was imaged using scanning electron microscopy (SEM) (Thermo-Fisher ApreoSEM). 

\subsubsection{Particle Zeta Potential Measurement using Dynamic Light Scattering}
Zeta potential was measured using a Malvern Zetasizer nano ZSP equipped with a He-Ne gas laser with a wavelength of 632.8 nm. A dilute dispersion with a concentration of 0.005 wt/v$\%$ of each cubic and spherical particles was prepared in Mill-Q water and sonicated for at least 10 minutes at 20 $^\circ$C temperature and 37 MHz frequency. Then, the suspension of particles was placed in the sample holder and allowed to equilibrate at 25 $^\circ$C  temperature, before the zeta potential was measured. For each particle, the reported zeta potential is an average of three separate measurements.

\subsubsection{Sample Preparation for Self Assembly and Microscopy Imaging}
First, the concentration of particles was adjusted to 4.55$\times$ $10^9$ particles ml$^{-1}$ (for 1.00±0.05 $\mu$m, 0.97±0.05 $\mu$m, 2.00±0.05 $\mu$m, and 2.06 ±0.05 $\mu$m spheres), 1.13$\times$ $10^9$ particles ml$^{-1}$ (for 4.55±0.05 $\mu$m, and 4.62 ±0.05 $\mu$m spheres) and 4.55$\times$ $10^7$ particles ml$^{-1}$ (for 0.83±0.03 $\mu$m, and 1.08±0.04 $\mu$m cubes). Before assembly, the particle dispersions were sonicated for at least 15 min at 20 $^\circ$C temperature and 37 MHz frequency to break up any aggregates. We mixed 200 $\mu$l of cubic particles with 200 $\mu$l spherical particles. We used a particle number ratio of spheres to cubes of 100:1 for the preparation of AB$_6$ and AB$_4$ valence colloidal molecules, and 25:1 for the preparation of AB$_2$ valence colloidal molecules. The assembled colloidal molecules were imaged using a custom-made sample holder equipped with a standard round glass coverslip on bottom and top. The prepared colloidal molecules were imaged using a Nikon TI-E A1 inverted bright field microscope using a 100x Oil objective (N.A 1.4) equipped with a Prime BSI Express camera (Teledyne Photometrics).

\subsubsection{Preparation of Fused Colloidal Molecules }
Colloidal molecules were separated from the excess spheres using a handheld magnet. Separated polystyrene-hematite colloidal molecules were fused as per the procedure described by Hueckel et al.\cite{hueckel2018mix} For this, 1 ml of 10 v/v$\%$ THF solution in Milli-Q water was prepared, and 50 $\mu$l of the separated colloidal molecules dispersion were added to the THF solution. In THF solution, the solid polymer spheres flow, fuse and stick to the cubes.  We allowed the fused colloidal molecules to settle under gravity, before removing the supernatant and replacing it with Milli-Q water to quench the process and re-solidify the spheres. The polystyrene spheres became firmly attached to the cubes through this procedure. For imaging, 20 $\mu$l of fused colloidal molecules were dried on a glass substrate at ambient temperature. Then, the substrate with dried colloidal molecules was sputter-coated with a 5nm platinum layer (Cressington 208HR sputter-coater) and imaged using scanning electron microscopy (SEM) (Thermo-Fisher ApreoSEM).

\subsection{Numerical Section}

\subsubsection{Monte Carlo Simulations}

To shed light on the self-assembly process of  satellite spheres  onto  cubes, we performed $NVT$ Monte Carlo simulations. We fixed a cube with side length $\sigma_c$ in the center of the simulation box and   started our simulations by randomly placing $N$ spheres of size $\sigma_s$ in the simulation box, thereby avoiding any overlap between the spheres and the cube. We varied the size ratio  $\alpha=\sigma_s/\sigma_c$  from $1.0$ to $5.6$. 
We set the number of spheres $N$ equal to either 25 or 100, so that the ratio between the number of spheres and  cubes matched that in the experiments. To account for the homogeneous surface charge of the cubes, we tessellated the cube surface with spherical beads of size $\sigma$. We used $\sigma$ as our unit of length in the simulations and fixed the side of the cube $\sigma_c = 5 \sigma$. To describe the electrostatic and excluded-volume interactions between the hematite cores and the satellite spheres in the experiments, we described the interactions between the spherical beads composing the cube and the satellite spheres by a hard-sphere (HS) and attractive screened Coulomb potential~\cite{dijkstra2002phase} $U_{Y,cs}(r_{ij})$  so that the total interaction potential $U_{cs}(r_{ij})$ reads
\begin{equation}
	\beta U_{cs}(r_{ij})=\begin{cases}
		\infty & r_{ij}\leq\sigma_{cs} \\
		\beta U_{Y,cs}(r_{ij}) & r_{ij}>\sigma_{cs}, 
	\end{cases}
\end{equation}
where $\beta=1/k_BT$ with $k_B$ the Boltzmann constant and $T$ the temperature, $\sigma_{cs}=(\sigma_{s}+\sigma)/2$, $r_{ij}$ denotes the center-of-mass distance between  bead $i$ and  satellite particle $j$, and 
\begin{equation}
	\beta U_{Y,cs}(r_{ij})=-\epsilon_{cs}\frac{\exp[-\kappa\sigma_{cs}(r_{ij}/\sigma_{cs}-1)]}{r_{ij}/\sigma_{cs}},
\end{equation}
where  the dimensionless prefactor $\epsilon_{cs}$ describes the interaction strength and $\kappa$ the inverse screening length.
The satellite spheres repel each other via a HS potential and a screened Coulomb repulsion $U_{Y,s}(r_{ij})$, and the total interaction potential $U_{ss}(r_{ij})$ reads
\begin{equation}
	\beta U_{ss}(r_{ij})=\begin{cases}
		\infty & r_{ij}\leq\sigma_{s} \\
		\beta U_{Y,s}(r_{ij}) & r_{ij}>\sigma_{s}, 
	\end{cases}
\end{equation}
with
\begin{equation}
	\beta U_{Y,s}(r_{ij})=\epsilon_{s}\frac{\exp[-\kappa\sigma_{s}(r_{ij}/\sigma_{s}-1)]}{r_{ij}/\sigma_{s}}. 
\end{equation}
As the experiments were performed in water, we set the Debye screening length  to  $0.1$ nm$^{-1}$, which corresponds to about $20 \sigma^{-1}$. We fixed the interaction strengths $\epsilon_s=10$ and  $10 \leq \epsilon_{cs} \leq 100$. The effect of varying $\epsilon_{cs}$ is shown in Figure S7, while in the main text we report results for $\epsilon_{cs}=15$. The resulting interaction potentials are extremely short-ranged and steep, and hence the effect on the assembly is that in the case that a sphere interacts with a cube, the sphere remains attached and only slightly fluctuates around their initial contact point. To study the self-assembly process, we performed  Monte Carlo  simulations for each size ratio for $10^6$ Monte Carlo steps starting from $100$ to $1000$ different initial configurations.

\subsubsection{Parking Problem}

We compared our particle clusters as obtained from  Monte Carlo simulations  with two different implementations of the ``random parking'' algorithm.~\cite{mansfield1996random} The random parking problem addresses the question of how many particles can be parked onto the surface of a central particle in case the sites are randomly chosen and the particles are non-overlapping. The random parking algorithm has been applied for parking spheres on a sphere.~\cite{schade2013tetrahedral,mansfield1996random} 

Here, we examined the random parking of spheres onto the surface of a cube. We considered two different implementations of the random parking algorithm. In the first case, $N$ spherical particles diffuse in the simulation box and a sphere is considered as ``parked'' onto the cube once the minimum distance condition between a sphere and a cube is satisfied. The minimum distance between the surface of a cube with  side length $\sigma_c$ and the center of a sphere with diameter $\sigma_s$ is equal to $\sigma_s/2$. To this end, we employed the Oriented Cuboid Sphere Intersection (OCSI) algorithm to detect overlap between cuboids and spheres.\cite{tonti2021fast} We refer to this implementation of the random parking algorithm as the parking problem based on  \textit{random adsorption} of satellite spheres onto a cube. The second implementation provides a simpler approach to the problem. We randomly selected a point on the surface of the cube and in case the no-overlap condition is satisfied, we placed a particle of diameter $\sigma_s$ at that position. This second implementation is referred to as the parking algorithm based on \textit{selecting a random site} on the surface of the cube. In both cases, we ignored the screened Coulomb interactions between the particles and only considered excluded-volume interactions. We investigated the particle clusters using the two different implementations of the random parking algorithm for the same range of size ratios $\alpha$  as in the Monte Carlo simulations. 

\section{Results and Discussion}
\subsection{Assembly and Purification Strategy for Colloidal Clusters}
To assemble colloidal molecules, we follow a procedure based on hetero-coagulation of oppositely charged colloids into small clusters. Upon mixing two species that attract each other, finite sized clusters are being formed if one species is used in excess of the other.~\cite{schade2013tetrahedral, tagliazucchi_kinetically_2014, van_ravensteijn_colloids_2014, demirors2015opposite} The particles that are being used in excess assemble on the second particle type until it is fully covered. This approach is not restricted to colloids interacting through electrostatic interactions, but it has also been shown for spheres modified with complementary DNA strands.~\cite{soto2002controlled, schade2013tetrahedral, chakraborty2022self} Here, we use cubic particles made from hematite as the core particles of the cluster because they feature two crucial advantages: (1) their non-spherical, cubic shape can be exploited as a mean to control the number of bound satellite particles and thus the shape of the resulting colloidal molecules, and (2) their permanent magnetic dipole moment can be used to separate the colloidal molecules from other particles after assembly. 

\begin{figure}[t!]
	\centering
	\includegraphics [width=0.99\textwidth]{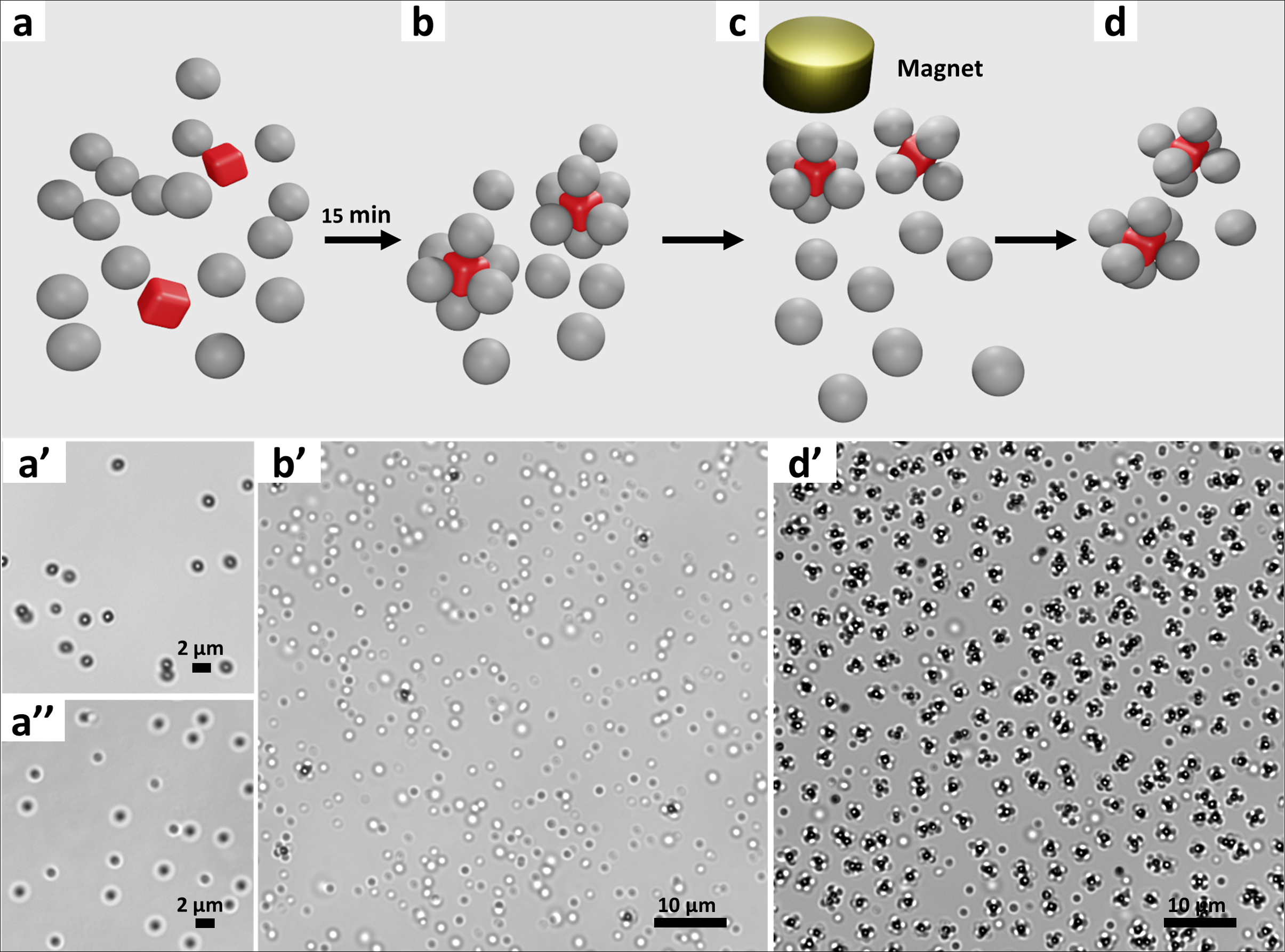} 
	\caption{Schematic illustrating the process of assembly and separation of colloidal molecules prepared from a binary mixture of negatively charged spheres  and positively charged cubes. (a) A mixture of positively charged cubes (depicted in red, and shown in a bright-field microscopy image in panel a')) and an excess number of negatively charged spheres (depicted in grey, and shown in the bright-field microscopy image in panel a")). (b) Spontaneously self-assembled colloidal molecules driven by electrostatic attractions between cubes and spheres. The excess number of spheres leads to saturation of the cube surface and hence finite size clusters. The cubic shape and size ratio of the cubes and spheres determine the cluster size. An example of the assembled clusters and the large excess amount of spheres is shown in panel (b'). (c) The permanent magnetic dipole of the cubic particles allows separation of colloidal molecules from unbound spheres with a magnet, yielding (d, d') isolated colloidal molecules after removal of suspended excess spheres.}
	\label{preparation}
\end{figure}

An overview of the general procedure for preparing and isolating colloidal molecules is shown in Figure~\ref{preparation}. In a typical experiment, we mixed positively charged cubes that were dispersed in water with an excess number (25:1 or 100:1, see Methods) of negatively charged spheres, either made from polystyrene or silica, of various sizes (see Table \ref{zeta}) and allowed them to self-assemble for 15 minutes. A schematic of the process is depicted in Figure~\ref{preparation}a and bright-field microscopy images of the initial particle dispersions are shown in panels a' and a". We found that longer assembly times did not significantly affect the number of bound satellite spheres. The excess of negatively charged spheres leads to quick saturation of the cubic core particles and hence formation of finite size clusters, the colloidal molecules.~\cite{soto2002controlled, schade2013tetrahedral} In each colloidal molecule a cube is surrounded by $N$ spherical particles, see Figure~\ref{preparation}b and b'. Their negative charge stabilizes the clusters against further aggregation, although this is much less pronounced than for clusters assembled in non-aqueous media.~\cite{demirors2015opposite} The assembled colloidal molecules can therefore be characterized by the number of adsorbed spheres $N$. The positions of the spheres on the cube particles are static due to the strong attraction between the particles. 

The permanent magnetic dipole of the hematite cores allows us to selectively isolate colloidal molecules with a magnetic field created using a handheld magnet, see Figure~\ref{preparation}c. Under the influence of the magnetic field, the colloidal molecules migrate towards the magnet, leaving behind a large number of excess spheres suspended in the upper layers of the fluid (see also discussion in Section 2.3). These could be removed by extracting the supernatant using a micropipette, and refilling the sample to its initial volume with Milli-Q water. To remove as many free spheres as possible, the same procedure was repeated three times, yielding high concentrations of colloidal molecules, see Figure~\ref{preparation}d and d'. 

\subsection{Geometric Effects in Determining the Cluster Size of Electrostatically Assembled Colloidal Molecules}

The main organizational principle in hetero-coagulation-type assembly of colloidal clusters is the geometry of the constituents and, potentially, electrostatic effects. Considering geometry only, the size ratio $\alpha$ between the satellite particles and the core has been found to determine their maximum number, for clusters made from two species of spheres.~\cite{miracle_influence_2003, schade2013tetrahedral} In addition, electrostatic repulsion between the satellite particles may lower their geometrically determined maximum number.~\cite{demirors2015opposite} The shape of the central particle may further limit the number of adsorbed spheres.~\cite{liu2021assembly}

To test the effects of electrostatic interactions and shape of the central core, we exploited the predictive power of numerical simulations to investigate their combined influence on the size and yield of the assembled clusters for different size ratios. We run Monte Carlo simulations in which a repulsive and an attractive Yukawa pair potential account for the screened Coulomb interactions between spherical satellite particles of diameter $\sigma_s$, and between the latter and the cubic core with edge length $\sigma_c$, respectively. As all experiments were performed in aqueous environments, we set the Debye screening length to $0.1$ nm$^{-1}$. The resulting Yukawa potential is thus extremely short-ranged and steep for both the attractive and repulsive interactions. We performed Monte Carlo simulations for different initial configurations in the presence of an excess of colloidal spheres. We provide more details on the numerical simulations in the Methods section. 

\begin{figure}[t!]
	\centering	\includegraphics [width=0.55\textwidth] {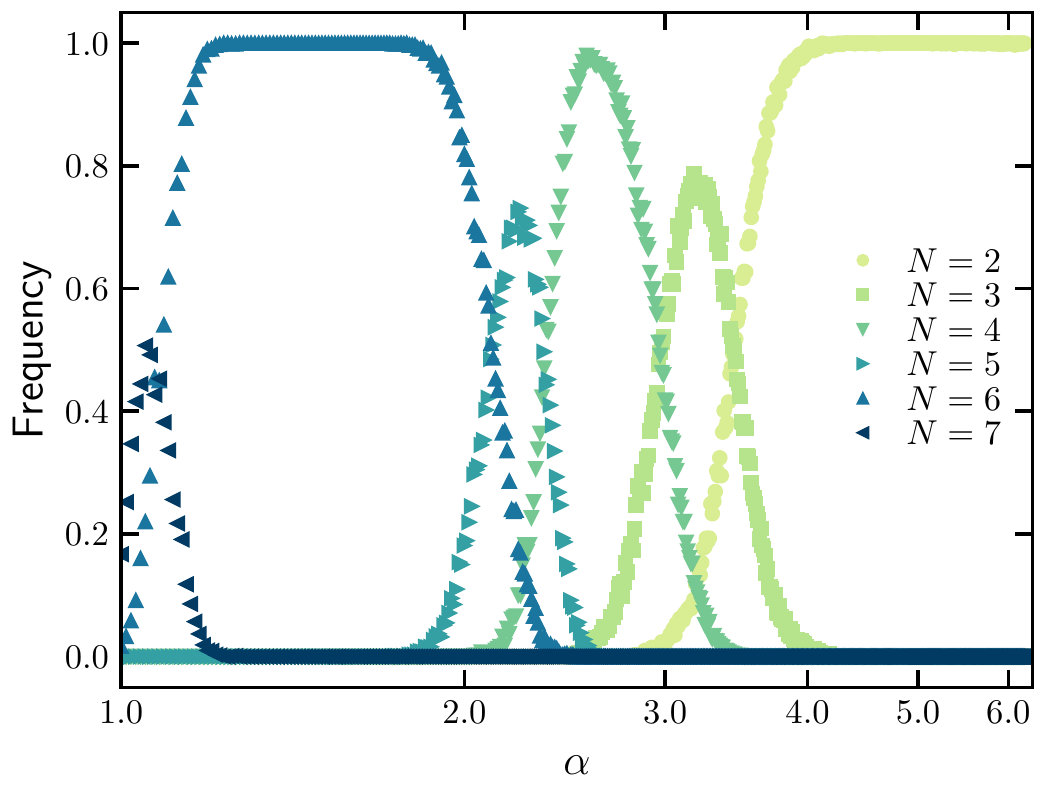}
	\caption{Normalized yield of clusters with $N$ adsorbed satellite spheres as a function of size ratio $\alpha=\sigma_s/\sigma_c$ as obtained from  Monte Carlo simulations with a repulsive screened Coulomb potential between the satellite spheres with diameter $\sigma_s$ and an attractive screened Coulomb potential between the satellite spheres   and the central cube with an edge length $\sigma_c$. 
	}
	\label{fig:avgclusterandyield_sim}
\end{figure}
We then determined the number of satellite spheres $N$ adsorbed onto a central cubic particle for varying  size ratios $\alpha=\sigma_s/\sigma_c$. In Figure~\ref{fig:avgclusterandyield_sim}, we plot the probability, i.e. the normalized yield, of clusters  with $N$  satellite spheres adsorbed onto the surface of a cube as a function of  size ratio $\alpha$. It emerges immediately that for $1.2 \lesssim \alpha \lesssim 2$ the cubic shape of the central core is the limiting factor for the number of bound satellite particles, allowing for only one particle per facet to be bound resulting into six-satellite-particle clusters. This is very different from the behavior that would be observed when satellite spheres are adsorbed onto the surface of a spherical central colloid, as was demonstrated in previous work.~\cite{schade2013tetrahedral} In the case of adsorption onto  a sphere, the range in which six-satellite-particle clusters are observed is much narrower and the corresponding maximum theoretical yield is  $30\%$ lower. The robustness of the resulting colloidal molecules towards variation in the size ratio also explains why colloidal molecules based on templating polyhedral particles could be achieved without an extensive search for optimal size ratios.~\cite{liu2021assembly}

In the case of adsorption onto a cube, the available space for binding a second sphere to a facet is greatly reduced when a sphere is already attached to that facet. 
For $\alpha \lesssim 1.2$, the anisotropic shape of the central particle has no or little effect on the number and organization of  bound particles.~\cite{hueckel2018mix} For $\alpha \gtrsim 2$, the percentage of clusters with less than six satellite particles gradually increases with $\alpha$, as the adsorption of spheres on an adjacent facet becomes increasingly restricted. It appears that the highest yield for clusters with four satellite particles is  reached for $\alpha=2.48$, which is slightly higher  than for the adsorption onto a sphere, where the highest yield was found for $\alpha=2.41$ for tetrahedrally arranged particles.
For an even higher size ratio, the number of satellite particles is further reduced up to $\alpha \approx 4$ for which their size is such that at most two particles can be bounded. 

We experimentally tested these predictions by assembling colloidal clusters from hematite cubes with edge lengths 0.83±0.03 $\mu$m (HC1) and 1.08±0.04 $\mu$m (HC2), and spheres made of either polystyrene or silica, with various sizes. In particular, we chose polystyrene spheres (PS) with a diameter of 1.00±0.05  $\mu$m (PS1), 2.00±0.05  $\mu$m (PS2) and 4.55±0.05  $\mu$m (PS3), and silica spheres (Sil) with size 0.97±0.05 $\mu$m (Sil1), 2.06±0.05 $\mu$m (Sil2) and 4.62±0.05 $\mu$m (Sil3). The size and charge of the various satellite and core particles used in this study are listed in Table~\ref{zeta}. The use of both polystyrene and silica spheres of similar sizes allows us to demonstrate the method's generality and identify the impact of electrostatic and gravitational effects on the cluster size distribution.
After assembly, we measured the cluster size distribution by manually counting a minimum of 100 clusters from at least two distinct microscope images or videos of magnetically separated clusters. We excluded unbound spheres and non-specific aggregates of the same type of particle. Videos were used to analyze three-dimensional clusters from small spheres and cubes ($\alpha$ =1.20). Still images were instead used for colloidal molecules made from larger spheres because they could be more easily distinguishable.

\begin{table}[b!]
	\centering
	\caption{Colloidal particles used in the experiments and their materials, notation, sizes, zeta potential and standard deviation of the zeta potential as measured by dynamic light scattering. }
	\label{zeta}
	\begin{tabular}{ccccc}
		\hline
		Material                                  & Notation & \makecell{Diameter\\ ($\mu$m)} & \makecell{Zeta Potential \\ (mV)} & \makecell{Standard Deviation \\ (mV)} \\ 
		\hline
		\multirow{3}{*}{Polystyrene Spheres (PS)} & PS1      & 1.00±0.05          & -38              & 5.2                    \\
		& PS2      & 2.00±0.05          & -47              & 2.0                    \\
		& PS3      & 4.55±0.05           & -45              & 3.2                    \\
		\multirow{3}{*}{Silica Spheres (Sil)}      & Sil1      & 0.97±0.05           & -45             & 8.9                    \\
		& Sil2      & 2.06±0.05           & -57              & 1.1                    \\
		& Sil3      & 4.62±0.05           & -79              & 4.6                    \\
		\multirow{2}{*}{Hematite Cubes (HC)}      & HC1      & 0.83±0.03           & 31               & 1.8                    \\
		& HC2      & 1.08±0.04           & 27               & 3.3                   \\
		\hline
	\end{tabular}
\end{table}

\begin{figure}[t!]
	\centering
	\includegraphics [width=\textwidth] {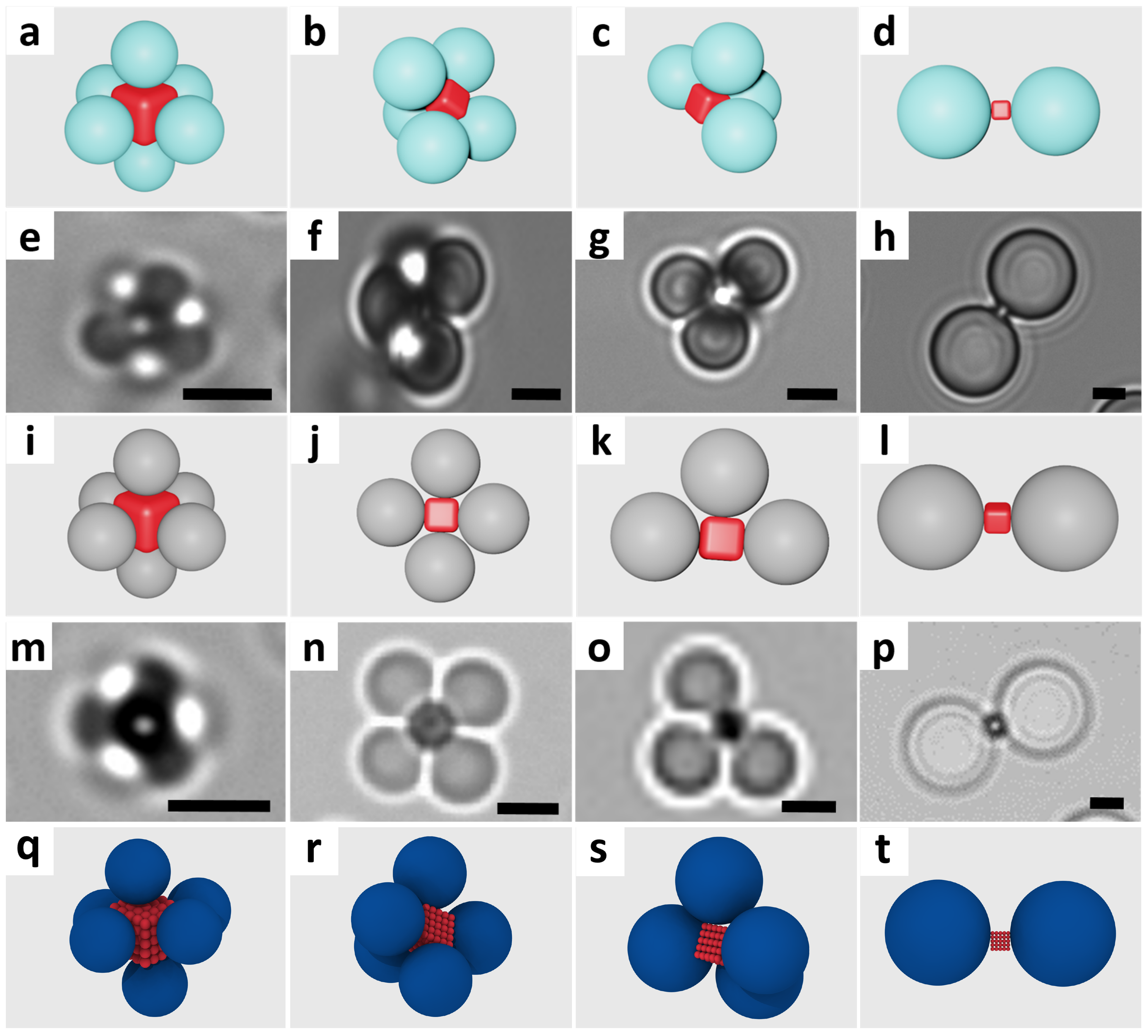}
	
	\caption{Electrostatically assembled colloidal molecules made from cubes and spheres for different size ratios $\alpha$. (a-d) and (i-l) rows show schematic depictions of the bright-field microscopy images of colloidal molecules prepared from (e-h) polystyrene spheres and hematite cubes, and (m-p) silica spheres and hematite cubes. For $\alpha > 1$, a maximum of six spheres can be bound to the central cube as shown for (e) $\alpha =1.2$ and (m) $\alpha=1.16$, and slightly larger size ratios (f) $\alpha=1.85$ and (n) $\alpha=1.90$. For $\alpha \gtrsim 2.40$, the number of bound particles decreases further, as experiments with (g) $\alpha=2.40$  and (o) $\alpha =2.48$ demonstrate. For $\alpha \gtrsim 4$, only two spheres can be bound per cube, see (h) $\alpha=5.48$, and (p) $\alpha=5.56$. Scale bar is 2 $\mu$m. (q-t) Representative snapshots of colloidal molecules obtained from Monte Carlo simulations for size ratios  $\alpha=1.20, 1.90, 2.40$, and $5.40$.}
	
	\label{models}
\end{figure}

\begin{figure*}[t!]
	\centering
	\includegraphics [width=0.99\textwidth] {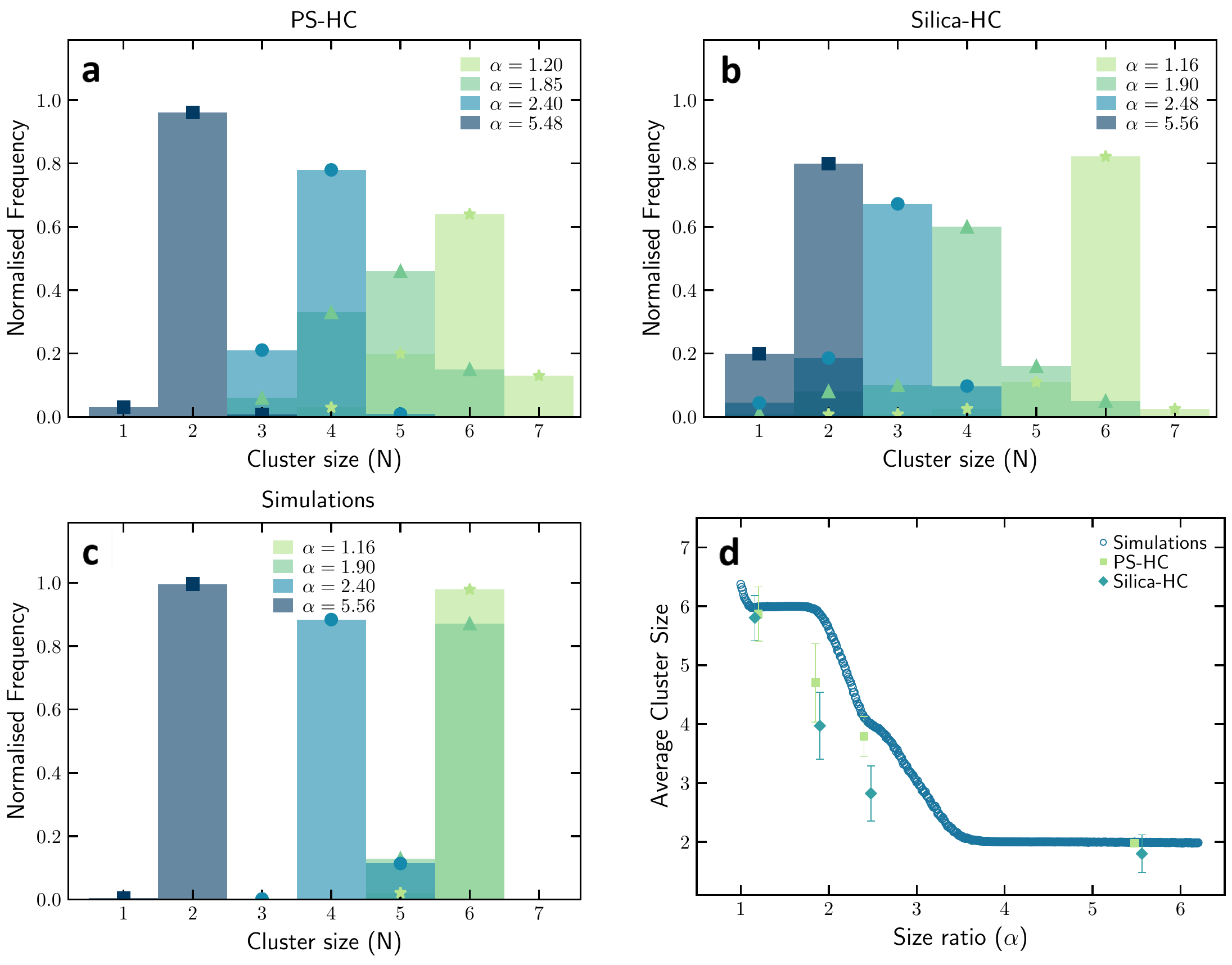}
	\caption{Cluster size distributions of colloidal molecules assembled from hematite cubes and (a) polystyrene or (b) silica spheres and (c) cluster size distributions for different values of the size ratio $\alpha$ obtained from Monte Carlo simulations. (d) Average cluster size of colloidal molecules assembled from polystyrene spheres and hematite cubes (PS-HC), silica spheres and hematite cubes (Silica-HC), and from Monte Carlo simulations as a function of size ratio $\alpha$. The standard deviation of the average cluster size was calculated from the average absolute deviation from the mean. Data in (c-d) correspond to $\epsilon_{cs}=15$.}
	\label{Cluster size}
\end{figure*}

AB$_6$-configuration clusters are indeed found for size ratios $\alpha=1.20$ (for polystyrene spheres) and $\alpha=1.16$ (for silica spheres), for which representative bright-field microscopy images together with a schematic drawing can be found in Figure~\ref{models}a, e, i and m. Additionally, we show a representative configuration from  Monte Carlo simulations for the same size ratios in Figure~\ref{models}q. Most importantly, such clusters constitute the most dominant fraction of clusters as compared to other configurations,
as shown by the cluster size distribution reported in Figure~\ref{Cluster size}a and b for polystyrene and silica spheres, respectively.
Consistently, at a comparable size ratio, this is also the most dominant fraction of clusters found in simulations, as reported in Figure~\ref{Cluster size}c.
The yield of 0.64 and 0.82 for PS and silica, respectively, at this size ratio is similar and higher than the maximum yield of about 0.7 that was found previously for AB$_6$ sphere-sphere clusters at the optimal size ratio, at which also AB$_5$ and AB$_7$ were found.~\cite{schade2013tetrahedral} We notice that this yield was only determined theoretically, and a presumably lower value can be expected for experimentally obtained sphere-sphere clusters. We further note that our experimentally chosen size ratio was not optimized for yield, although it is close to the geometrically expected size ratio for a maximum yield. 

By increasing the size ratio, the number of spheres bound per cube decreases, as expected. At $\alpha=1.85$ and $\alpha=1.90$, clusters with a predominantly AB$_5$ and AB$_4$ geometry were observed for polystyrene and silica spheres, respectively. At this size ratio, a higher yield of AB$_4$ of about 0.6 is achieved for silica spheres, with such clusters typically having all particles arranged in a plane. Clusters from both types of particles have a broader distribution, meaning that at this size ratio the transition from AB$_6$ to AB$_2$ is still ongoing. From the schematic drawing shown in Figure~\ref{models}b and the representative image from Monte Carlo simulations in Figure~\ref{models}r it becomes apparent that indeed the packing on the surface of the cube becomes tight at this size ratio. Close-ups of bright-field microscopy images of the corresponding clusters are shown in Figure~\ref{models}f and n, for polystyrene and silica spheres, respectively. 

For slightly larger polystyrene spheres with a size ratio $\alpha=2.40$, we find predominantly, i.e. in 78\% of all cases, clusters with AB$_4$ geometry (Figure~\ref{models}c and g, and Figure~\ref{Cluster size}a), while 21\% of the clusters feature three spheres. This size ratio is still below the one at which the transition AB$_2$ would be expected. From the bright-field images (Figure~\ref{models}g) and simulation snapshot (Figure~\ref{models}s) it is visible that the four outer spheres typically do not lie in the same plane, but feature three particles in the same plane and one particle above. 
At a similar, although slightly higher, size ratio of $\alpha=2.48$, clusters containing silica particles already feature mostly AB$_3$-type clusters, that is 67\%. In contrast to the polystyrene-based clusters, the three spheres and cube are arranged in a plane, see Figure~\ref{models}k, and o.

Finally, at an even higher size ratio, we find almost exclusively AB$_2$ clusters for both polystyrene ($\alpha=5.48$) and silica spheres ($\alpha=5.56$), with a yield of 0.96 for PS and 0.80 for silica based clusters, as shown in Figure \ref{Cluster size}. A schematic drawing and corresponding bright-field microscopy images of the resulting clusters, illustrating the exclusion of a third particle from the surface of the cube, can be found in Figure~\ref{models}d, h, l, and p. Accordingly, in simulations, at $\alpha=5.40$ we only find clusters like the one shown in Figure~\ref{models}t, as demonstrated by the normalized frequency reported in Figure~\ref{Cluster size}c.

Comparing the cluster size distributions shown in Figure~\ref{Cluster size}a and b, we find that at similar size ratios, polystyrene based clusters have a slightly larger cluster size, i.e. more particles are bound to the  surface of the cube. This becomes even clearer in Figure~\ref{Cluster size}d, where the average cluster size for both polystyrene and silica spheres is shown. Again, on average the cluster size of polystyrene-hematite colloidal molecules is slightly larger than that of silica-hematite ones. In addition, for a similar size ratio, we noticed a wider cluster size distribution in polystyrene based colloidal molecules. The consistently larger size of the assembled colloidal molecules obtained from using polystyrene spheres can be understood from the combined effect of electrostatic repulsion and gravitational confinement to quasi-2D. 
Silica particles are more negatively charged than polystyrene particles of the same size, as demonstrated by the zeta potential measurements displayed in Table~\ref{zeta}. A stronger repulsion implies that fewer silica particles can be accommodated on the cube's surface, equivalent to an effectively larger size. Electrostatic repulsion between the outer spheres have also been found to play a role in experiments and simulations on sphere-sphere clusters in a non-aqueous solvent.~\cite{demirors2015opposite} There, it was demonstrated that the cluster size of colloidal molecules is dependent on the size ratio, charge, and density of the particles, and in particular that an increase in particle charge implies a decrease in the average size of the colloidal molecules. However, we expect that the impact of the different zeta potential values on the cluster size in our case is still small compared to the geometric effects imposed by the size ratio of the cube and sphere, and the anisotropic shape of the cube. 

Secondly, the higher density of silica leads to a lower gravitational height compared to that of polystyrene spheres, whose density is close to that of the surrounding medium. While this effect is negligible for small silica spheres, it is clearly visible for larger ones, where quasi-2D clusters are being formed. At size ratios that should still allow binding of six spheres to the cube, often only four are being found. This implies that despite tumbling the sample during assembly which should enable 3D clusters, the clusters reorganize with spheres breaking off due to the quasi-2D confinement in the sample holder. The uniformity of the assembled colloidal molecules exemplifies that gravitational confinement can be used as an additional strategy to assemble AB$_4$ clusters instead of AB$_6$ with a rather uniform shape.

When comparing the size distribution of binary clusters assembled from spheres and cubes with those made previously from spheres only, we find that the yields for a specific AB$_N$ are higher than what has been achieved for spheres,~\cite{soto2002controlled, schade2013tetrahedral, demirors2015opposite} with the exception of AB$_4$ at the very specific size ratio $\alpha=2.41$, which constitutes a special case. In addition, the highest yield for AB$_4$ type colloidal molecules with a cube at the center is slightly shifted to higher $\alpha$,~\cite{schade2013tetrahedral} just like other specific cluster sizes. This is a direct consequence of the faceting of the cubic template.

We also note that for all size ratios, we consistently find smaller yields and slightly wider distributions of the most likely cluster size in experiments compared to the ones predicted by Monte Carlo simulations, see Figure~\ref{Cluster size}.  We hypothesize that this discrepancy stems from a combination of the polydispersity in particle size and shape in the experiments for both satellite and core particles. Furthermore, the experimental cluster size distributions were obtained after magnetic separation of the colloidal molecules, a procedure that might break off some of the adsorbed spheres. This is corroborated by the fact that we never observed larger cluster sizes in experiments than those theoretically predicted by simulations. It is worth noting that on average, the sizes of the experimentally and numerically observed clusters are always comparable, with a remarkably good agreement in particular for AB$_6$ and AB$_2$ clusters. Such comparison is reported in Figure~\ref{Cluster size}d.

Finally, to obtain a deeper understanding on the assembly mechanism and on the obtained cluster size distributions, it is interesting to compare our outcomes to what was previously found for sphere-sphere clusters, assembled from two different species of spheres, one of which in excess with respect to the other. In particular, it was shown that the resulting cluster size distributions agree well with those found for the ``random parking'' problem. Therein, no interaction potential is taken into account and spheres are simply \textit{parked} onto the surface of the central colloid (see Methods), given that they stick randomly and irreversibly on each other surfaces.~\cite{schade2013tetrahedral}
We thus investigate  whether this also holds in the case of a cubic core particle. To this end, we determined the cluster size distributions and averaged cluster size by implementing two versions of the random parking algorithm. We first employed a random parking algorithm based on random adsorption of satellite spheres onto a cube. As shown in Figure S1-S3 of the Supplementary Material, we clearly observe that the number of spheres attached to a cube is higher than observed in experiments and Monte Carlo simulations, presented earlier in this section. Analyzing the resulting clusters as obtained from the random adsorption parking algorithm, we observe a tendency of satellite particles to be located on the edges and corners of the cubes rather than the faces, leading to more free space for other satellite particles to be bound and thus to higher averaged numbers of bound spheres. In Figure S3, we show typical configurations of the resulting clusters. 

We additionally employed a much simpler random parking algorithm based on selecting a random site on the surface of a cube. Remarkably, we find that the  cluster size distribution and average cluster size obtained with this approach matches very well with the ones obtained from experiments and  Monte Carlo simulations, as shown in Figure S4 and S5. As the probability to select a random site on one of the faces of the cube is much higher than on one of the edges or  corners of the cube, this finding confirms again the preference of the satellite spheres to be located on the faces of the cube. 
The different cluster size distributions as obtained from the random parking algorithm based on random site selection or random adsorption can be explained as follows. In the case of random adsorption, the adsorption site is determined by the minimal center-to-surface distance of the satellite sphere and the cubic core, resulting into a cuboidal surface for the center of masses of the bound satellite spheres. Hence, the  preference to stick to the corners and edges of the cube is much higher for random adsorption than in the case of selecting a random site on the surface of a cube, leading to significantly different cluster size distributions.

\subsection{Magnetic Separation of Colloidal Molecules and Fusion of the Adsorbed Spheres}

The permanent magnetic dipole moment of the central cube allows us to separate the colloidal molecules after their assembly from the unbound spheres. Easy isolation of the colloidal molecules is useful for any further experimental studies that rely on high concentrations and purity of large quantities. We demonstrate the principle for all samples investigated here, that is, both for different size ratios and sphere types.

\begin{figure}[b!]
	\centering
	\includegraphics [width=0.99\textwidth] {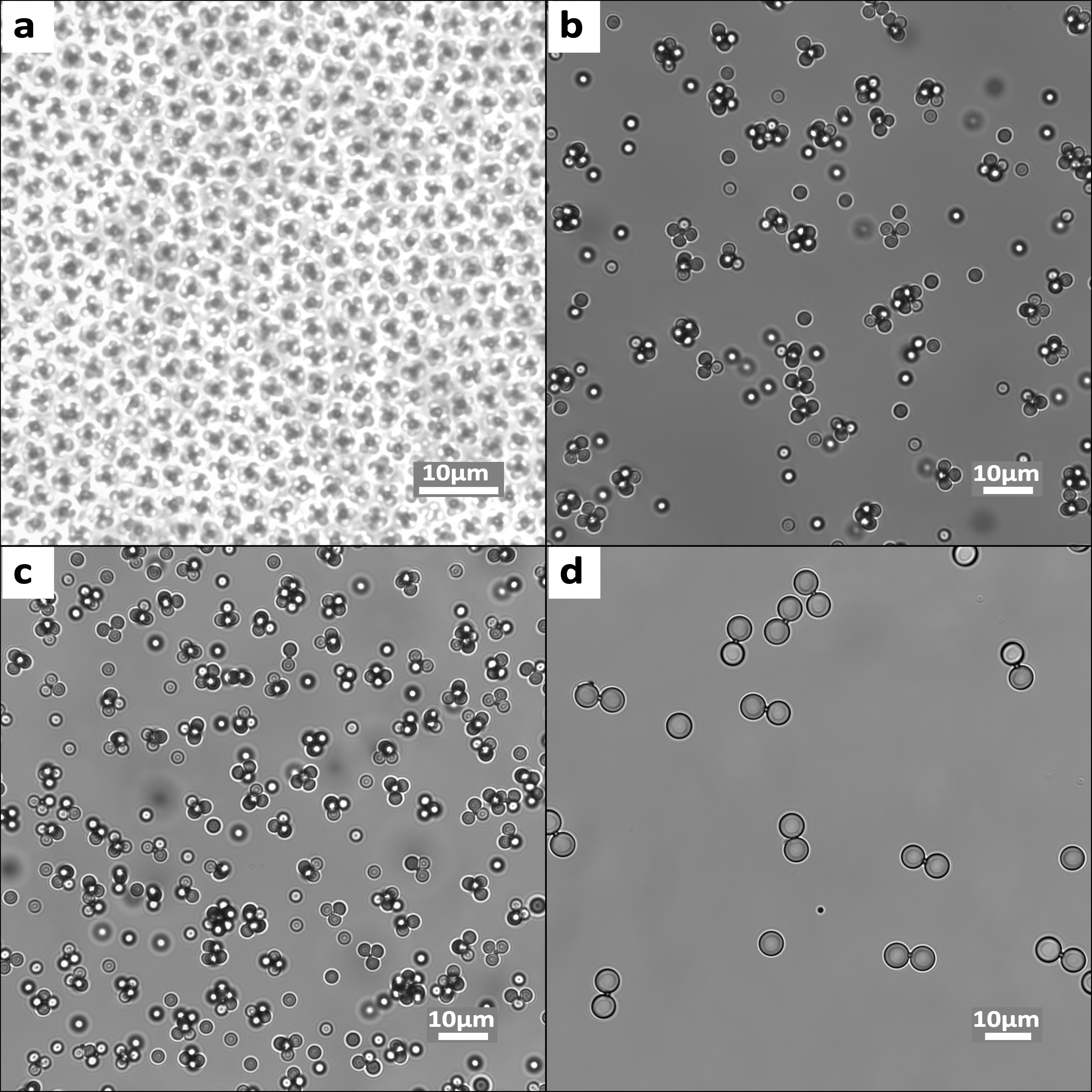} 
	\caption{Bright-field microscopy images of magnetically separated  colloidal molecules assembled from polystyrene (PS) and hematite cubes (HC) using different size ratios $\alpha$: (a) $\alpha$=1.20 assembled from PS1 and HC1, (b) $\alpha$=1.85 (PS2 and HC2), (c) $\alpha$=2.40 (PS2 and HC1), and (d) $\alpha$=5.48 (PS3 and HC1).}
	\label{PS}
\end{figure}

\begin{figure*}[t!]
	\centering
	\includegraphics[width=\textwidth]{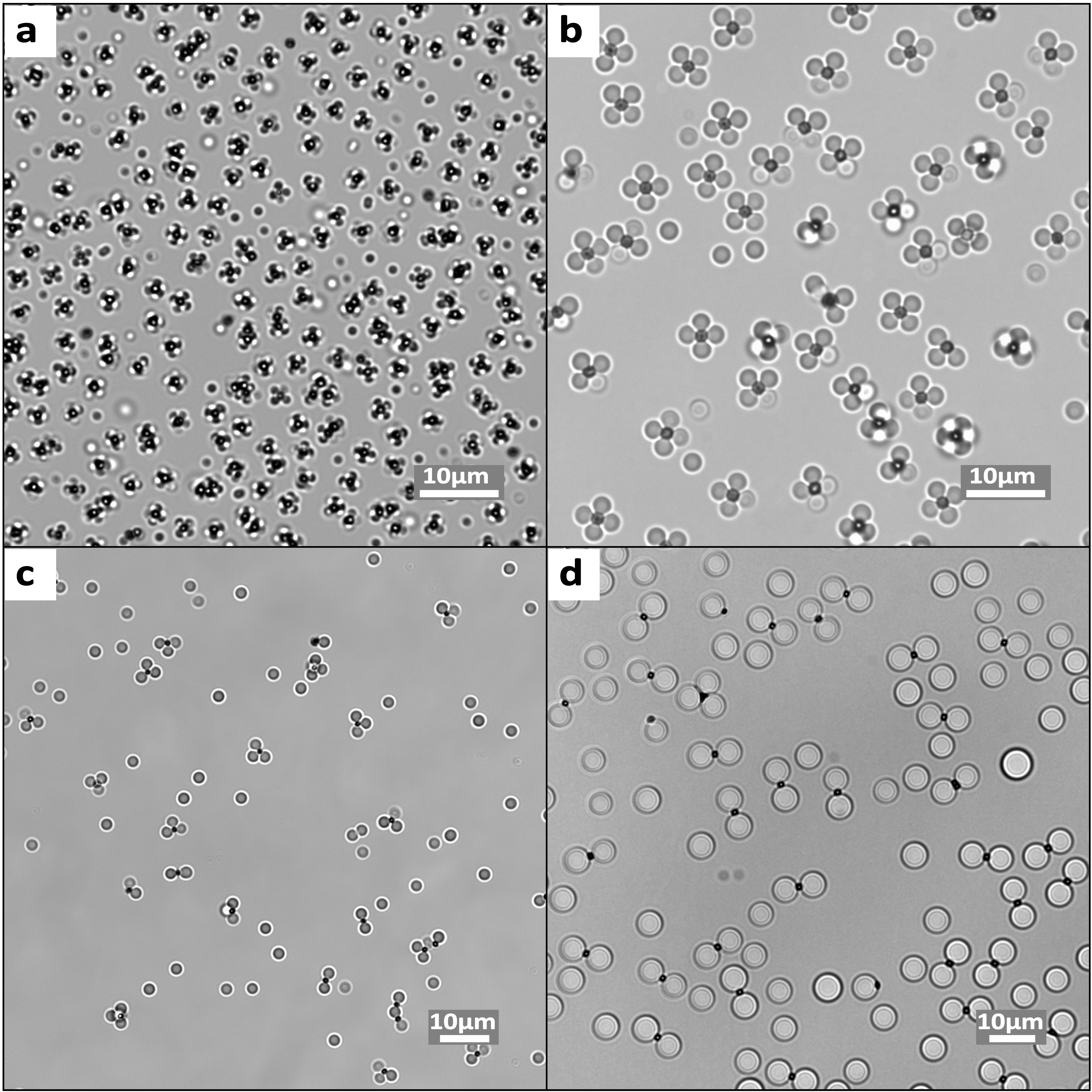} 
	\caption{Bright-field microscopy images of   colloidal molecules assembled from silica spheres (Sil) and hematite cubes (HC) using different size ratios $\alpha$: (a) $\alpha$ = 1.16 (assembled from Sil1 and HC1), (b) $\alpha$ =1.90 (Sil2 and HC2), (c) $\alpha$ =2.48 (Sil2 and HC1), and (d) $\alpha$ =5.56 (Sil3 an HC1).}
	\label{SS}
\end{figure*}

Figure~\ref{PS} shows representative light microscopy images of separated colloidal molecules obtained from different mixtures of polystyrene and hematite colloids. From the large field-of-view images it is evident how high yields of colloidal molecules can be obtained in this way. Such images were used to obtain the cluster size distributions reported in Figure~\ref{Cluster size}. For example, as discussed above, for $\alpha$ = 1.20 shown in Figure~\ref{PS}a, most colloidal molecules feature one sphere attached to each of the six faces of the cube, resulting in high yields of AB$_6$-type colloidal molecules. With increasing size ratio from Figure~\ref{PS}a to d, the number of bound spheres decreases. At size ratios far from the critical ones identified above, that is Figure~\ref{PS}b and c, the assembled colloidal molecules feature a wider range of number of bound particles. For $\alpha=5.48$, only AB$_2$ colloidal molecules are seen, with occasionally single spheres or cubes from a broken cluster.

To demonstrate the generality of the method, we also separated  colloidal molecules fabricated from silica spheres and hematite cubes. The representative light microscopy images of  separated Sil-HC colloidal molecules obtained for the four size ratios discussed above are shown in Figure~\ref{SS}a-d. The respective size distributions are shown in Figure~\ref{Cluster size}b. Again, the number of bound spheres per cube clearly reduces with  increasing size ratio. In addition, the larger size and higher density of the silica spheres that was employed in the experiments leads to a large fraction of quasi-2D colloidal molecules, which increases with the size of the silica spheres. For this reason, AB$_3$ and AB$_4$-type clusters prepared from silica spheres, as shown in Figure~\ref{SS}b and c, have a different geometry than those assembled from polystyrene, see Figure~\ref{PS}b and c.

\begin{figure}[t!]
	\centering
	\includegraphics [width=0.99\textwidth] {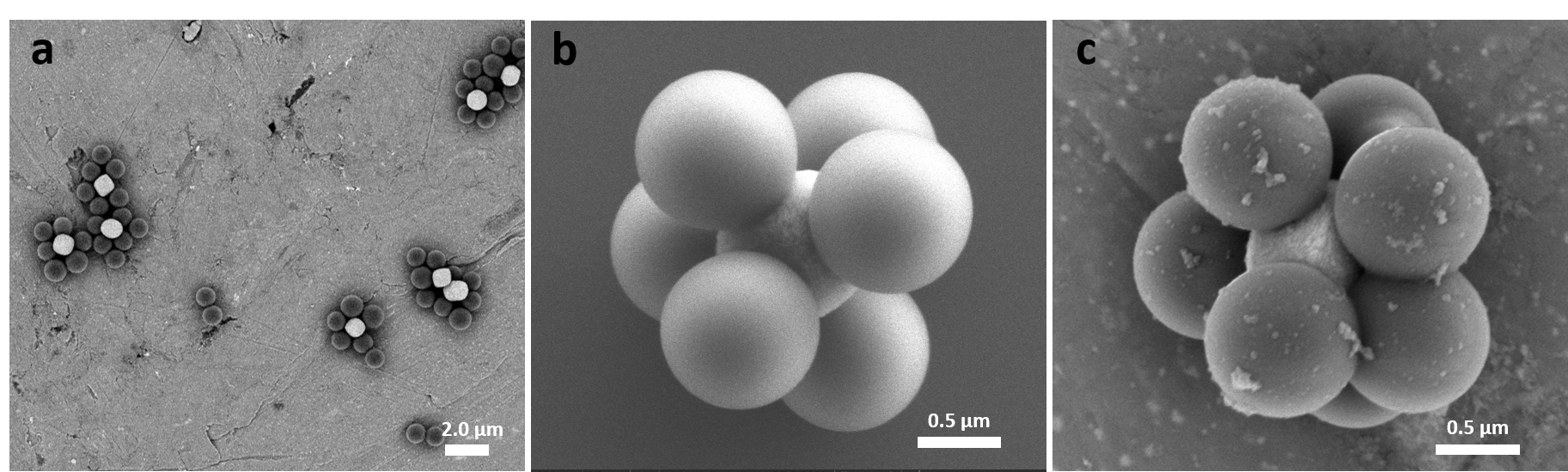} 
	\caption{Scanning electron microscopy images of colloidal molecules assembled from polystyrene spheres and hematite cubes. (a) Dried, unfused colloidal molecules, (b,c) Fused colloidal molecule with six and seven bound spheres, respectively.}
	\label{sem}
\end{figure}

A closer inspection of the sphere arrangement on the cubes is in principle possible by drying the colloidal molecules at room temperature to be able to image them using electron microscopy. However, solvent evaporation induced capillary forces disengaged the particles from the cubes, as shown in Figure \ref{sem}a. That clusters were so easily broken also may point out that some clusters might have broken during the magnetic separation. To address this issue, we used a previously described technique for merging colloidal molecules by the addition of a plasticizer.~\cite{hueckel2018mix} The results for representative fused AB$_6$ and AB$_7$ colloidal molecules are presented in Figure \ref{sem}b and c, and were captured using a scanning electron microscope. It is worth noting that, as also observed in simulations, not all spheres are perfectly centered on the cube face and that in the AB$_6$ cluster, each  face of the cube has only one sphere attached.

\section{Summary and Conclusions}
{In our work, we investigated the electrostatically-driven self-assembly of colloidal molecules, determining that simple ingredients such as particle shape and size ratio between colloidal particles are crucial for the clever control of their valence. Using positively charged colloidal cubes and negatively charged spheres in the assembly, we found that the six-fold geometry of the cube can provide AB$_6$, AB$_4$, and AB$_2$-type colloidal molecules in high yields. By complementing experiments with Monte Carlo simulations, we determined the range of size ratios between spheres and cubes for obtaining each colloidal molecule type, and we identified that electrostatic repulsion between the spheres is important to leverage the templating effect of the colloidal cube. At the same time, we assessed how such interactions allow the absorption of the spheres to be preferentially directed to the faces of the cube, thus discouraging the formation of irregular and asymmetric clusters due to the presence of satellite particles on the edges of the core. For this reason, we are able to describe the assembly process of colloidal molecules by a simple parking algorithm in which the adsorption points on the surface of the cube are selected randomly. 

The protocol for assembling colloidal molecules we presented is innovative in several respects. First of all, the control of the valence is simply achieved via geometrical factors. In fact, with a single templating shape, it is sufficient to choose the appropriate size of core and satellite particles for obtaining a desired coordination number. Compared with the use of a sphere as the central particle, ~\cite{demirors2015opposite, schade2013tetrahedral} we can achieve both greater selectivity in the type of molecule obtained as well as a higher yield, besides obtaining additional control over the relative position of absorption. Secondly, the presented process is mediated only by electrostatic interactions. Other procedures for obtaining colloidal molecules with different coordination numbers typically require complex functionalization through DNA,~\cite{schade2013tetrahedral} colloidal fusion.~\cite{gong2017patchy} In the latter case, for instance, the size ratio has never been used as a control parameter.~\cite{liu2021assembly} Exploiting electrostatic interactions for the assembly further makes our protocol highly versatile in terms of the choice of the material of the starting building blocks. In fact, we showed that similar colloidal molecules were obtained from polystyrene and silica particles at the same size ratios, matching the average cluster size prediction of Monte Carlo simulations. Finally, by choosing the material of the central core to be hematite, we took advantage of its magnetic properties to separate colloidal molecules from excess spherical particles,~\cite{ozaki1986reversible} thus allowing us to further increase the overall purity. For this reason, we could envision this method to be further exploited for the large-scale fabrication of such colloidal molecules. In contrast, the purification step in other assemblies typically involves multiple sedimentation-redispersion processes by in density gradient centrifugation~\cite{manoharan2003dense} or fluorescence-activated cell sorting.~\cite{mage2019shape}

We believe that the simplicity and robustness of our approach can be straightforwardly apllied to other template geometries beyond the cube, and therefore may be used to assemble colloidal molecules with a manifold of different shapes.~\cite{li2020colloidal} The electrostatic assembly method implies that colloids from various materials may be employed and combined into colloidal molecules. Potentially, the shape-based strategy might be extended to other driving mechanisms for self-assembly, such as depletion-induced~\cite{sacanna2010lock} or hydrophobic interaction.~\cite{chen2011supracolloidal,chen2011directed} The high yields obtained through this particle-shape driven method pave the way for using these particles in further self-assembly experiments towards materials with novel properties~\cite{glotzer2007anisotropy} or as model systems for real molecules.~\cite{verweij2020flexibility}}

\section*{CRediT authorship contribution statement}
\textbf{Yogesh Shelke:} Conceptualization, Methodology, Validation, Data curation, Formal analysis, Investigation, Visualization. \textbf{Susana Mar\'in-Aguilar:} Data curation, Methodology, Validation, Software, Formal analysis, Investigation, Visualization. \textbf{Fabrizio Camerin:} Data curation, Methodology, Validation, Formal analysis, Investigation, Visualization.
\textbf{Marjolein Dijkstra:} Methodology, Resources, Supervision, Funding acquisition.
\textbf{Daniela J. Kraft:} Methodology, Resources, Supervision, Funding acquisition. All authors equally contributed to writing the original draft, and reviewing and editing of the final draft.

\section*{Declaration of Competing Interest}
The authors declare that they have no known competing financial interests or personal relationships that could have appeared to influence the work reported in this paper.

\section*{Acknowledgements} 
The authors thank Rachel Doherty for SEM imaging. DJK gratefully acknowledges funding from the European Research Council (ERC Starting Grant number 758383, RECONFMAT). SMA, FC and MD acknowledge financial support from the European Research Council (ERC Advanced Grant number  ERC-2019-ADV-H2020 884902, SoftML).



\clearpage
\newpage

\begin{center}
\Large
Exploiting anisotropic particle shape to electrostatically assemble colloidal molecules with high yield and purity\\ \bigskip Supplementary Material

\normalsize
\bigskip
Yogesh Shelke\textsuperscript{ 1}, Susana Mar\'in-Aguilar\textsuperscript{ 2}, Fabrizio Camerin\textsuperscript{ 2}, Marjolein Dijkstra\textsuperscript{ 2}, Daniela J. Kraft\textsuperscript{ 1}\\
\medskip
\small
\textit{%
\textsuperscript{1}Soft Matter Physics, Huygens-Kamerlingh Onnes Laboratory, Leiden University, PO Box 9504, 2300 RA Leiden, The Netherlands\\
\textsuperscript{2}Soft Condensed Matter, Debye Institute for Nanomaterials Science, Utrecht University, Princetonplein 1, 3584 CC Utrecht, The Netherlands\\
}

%
%

\end{center}

\renewcommand{\thefigure}{S\arabic{figure}}\setcounter{figure}{0}

 \section*{Parking algorithm based on random adsorption}

 \begin{figure}[h!]
	\centering
	\includegraphics [width=0.6\textwidth]{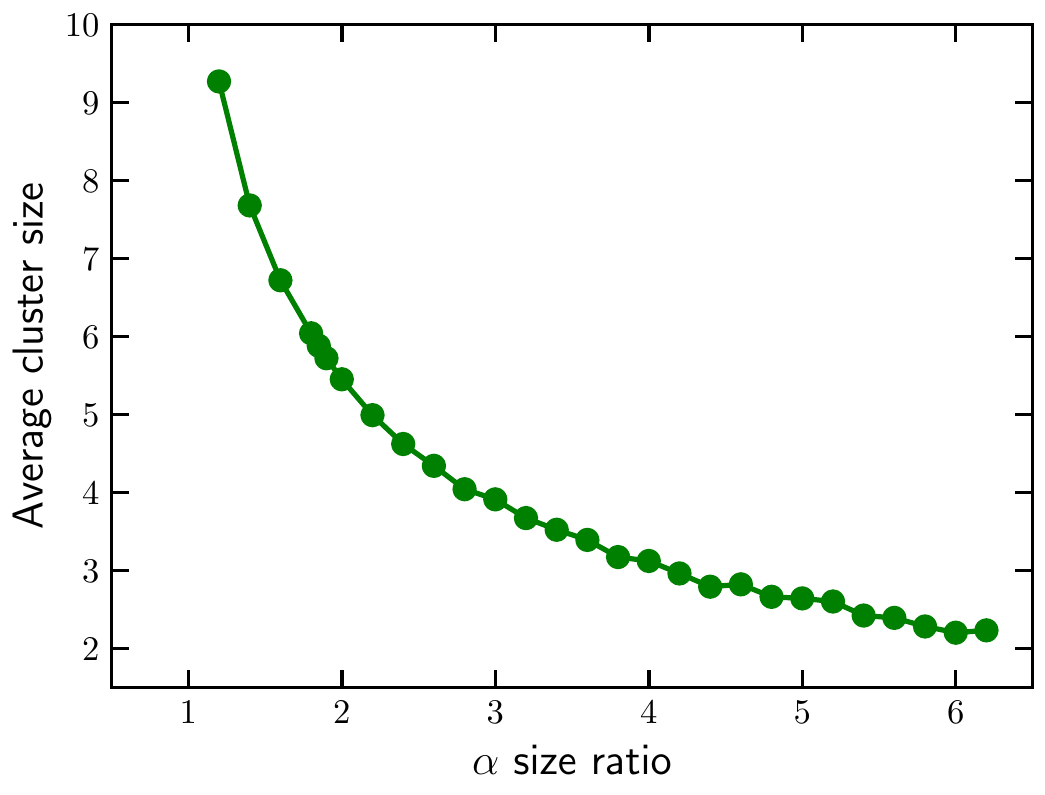} 
	\caption{Average cluster size  as a function of  size ratio $\alpha$ as obtained from the random parking algorithm based on random adsorption of spheres onto the surface of a cube.}
	\label{fig:1}
\end{figure}

\begin{figure}[t!]
	\centering
	\includegraphics [width=1\textwidth]{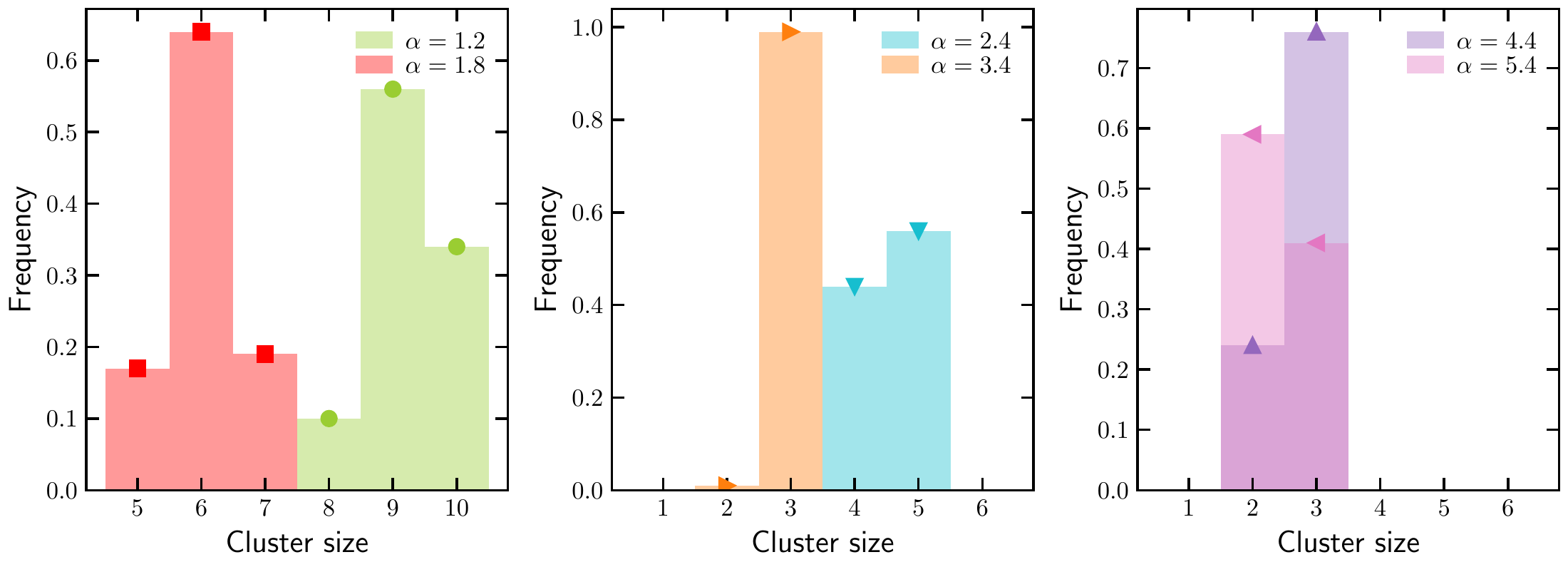} 
	\caption{Cluster size distributions as obtained from the random parking algorithm based on random adsorption of spheres  onto the surface of a cube  for different size ratios $\alpha$.}
	\label{fig:2}
\end{figure}

\begin{figure}
	\centering
	\includegraphics [width=0.99\textwidth]{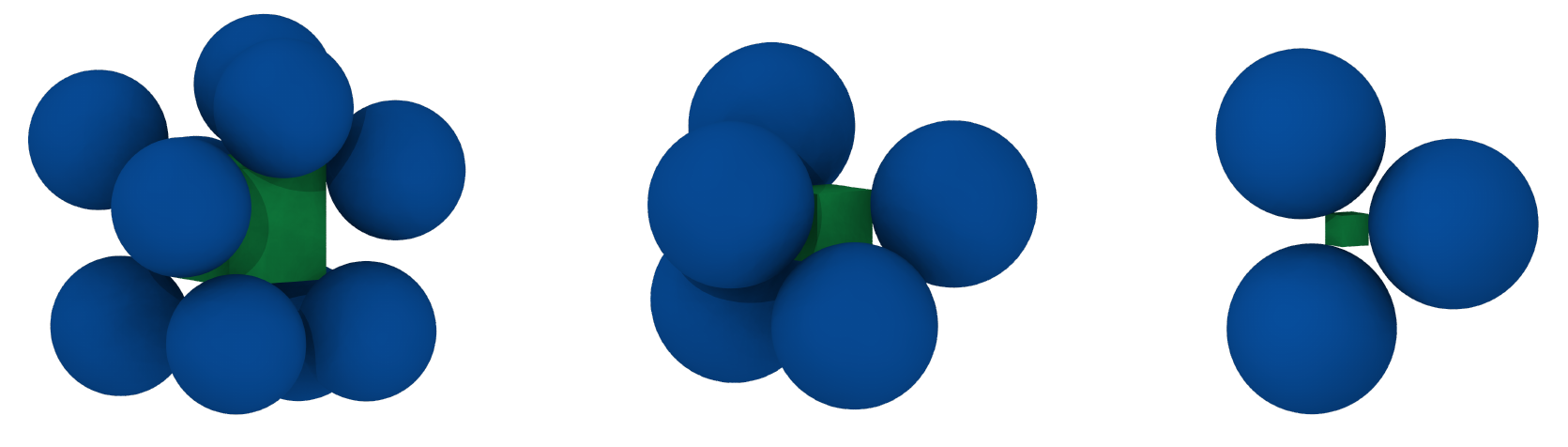} 
	\caption{Typical configurations as obtained from the random parking algorithm based on random adsorption of spheres onto the surface of a cube for different size ratios $\alpha=1.2, 2.4$ and 5.4.}
	\label{fig:3}
\end{figure}

In Figure~\ref{fig:1}, we show the average cluster size as a function of size ratio $\alpha=\sigma_s/\sigma_c$ for the random parking algorithm based on the random adsorption of spheres with diameter $\sigma_s$ onto the surface of a cube with edge length $\sigma_c$, as described in the Methods section. We note that the average cluster size, defined as the average number of particles attached to the cube, is higher than  in experiments and Monte Carlo simulations. Analyzing the resulting clusters as obtained from the random adsorption parking algorithm, we observe  a tendency of satellite particles to be adsorbed onto the edges and corners of the cubes rather than the faces,  leading to more free space for other satellite particles to be bounded  and thus to higher averaged numbers of bounded spheres. In addition, we show cluster size distributions in Figure~\ref{fig:2}, which appear to be  broader and shifted towards  higher numbers of bound spheres than the ones reported in the main text for  Monte Carlo simulations. Even the highest size ratio $\alpha$ shows a broader distribution, where not only clusters of two but also of three bound satellite spheres could be observed. In Figure~\ref{fig:3}, we display typical configurations as obtained using the random adsorption parking algorithm for varying size ratios.

\section*{Parking algorithm based on selecting a random site}
\begin{figure}[h]
	\centering
	\includegraphics [width=0.6\textwidth]{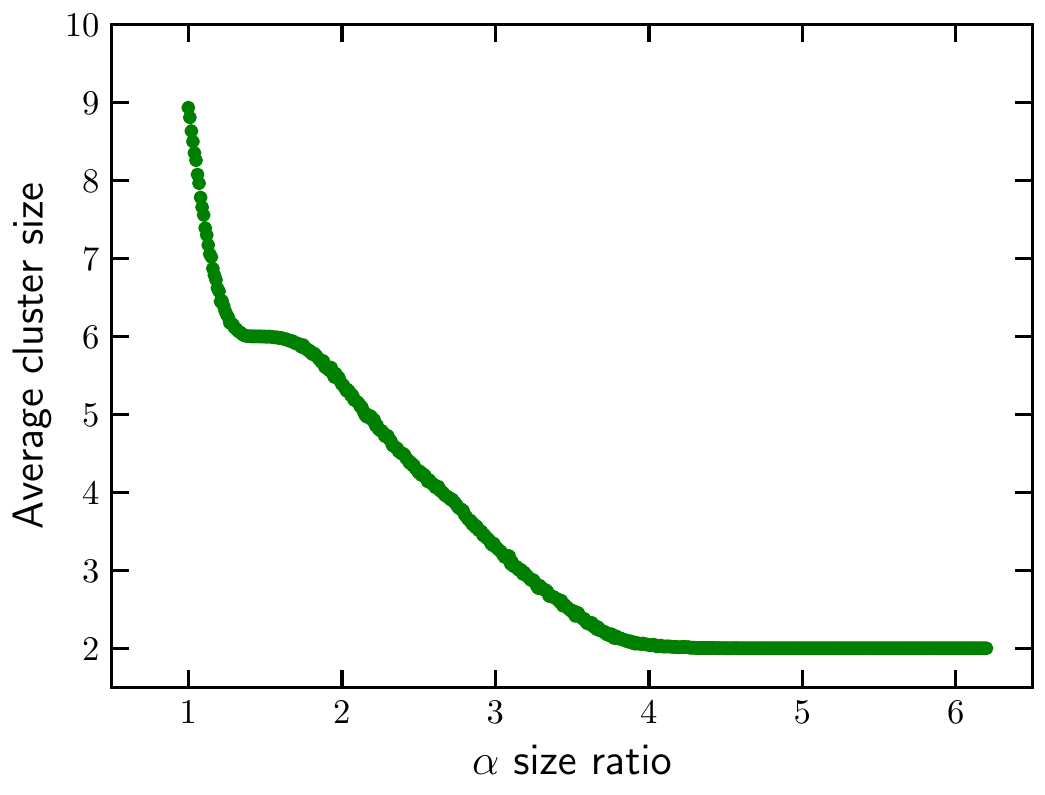} 
	\caption{Average cluster size  as a function of  size ratio $\alpha$ as obtained from the random parking algorithm based on selecting a random site  for spheres attached to the surface of a cube.}
	\label{fig:4}
\end{figure}

\begin{figure}[h]
	\centering
	\includegraphics [width=1\textwidth]{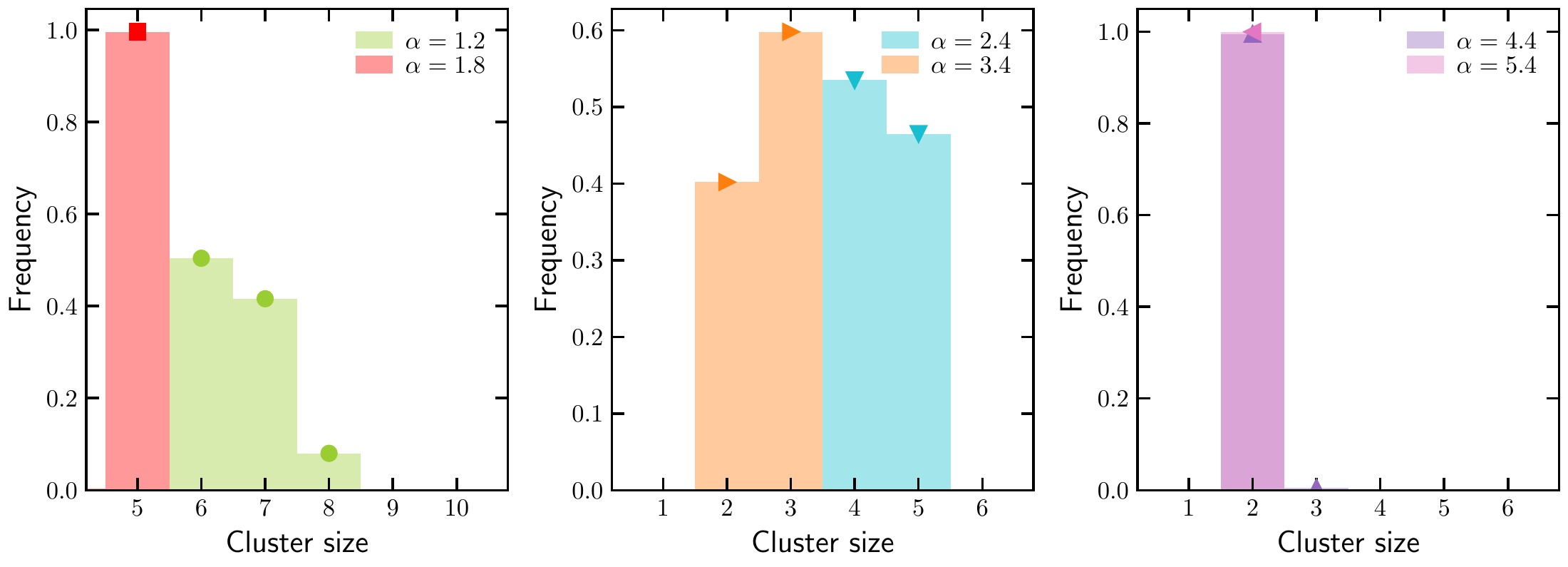} 
	\caption{Cluster size distributions   as obtained from the random parking algorithm based on selecting a random site for spheres attached to  the surface of a cube  for different size ratios $\alpha$. }
	\label{fig:5}
\end{figure}

\begin{figure}[h]
	\centering
	\includegraphics [width=0.99\textwidth]{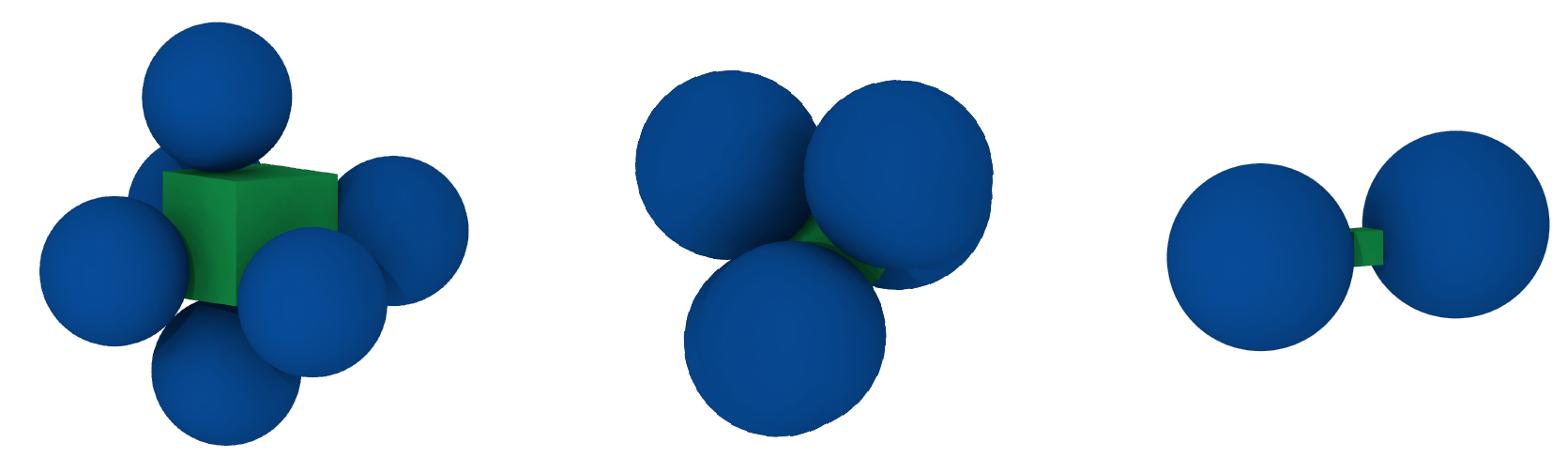} 
	\caption{Typical configurations as obtained from the random parking algorithm based on selecting a random site for spheres attached to  the surface of a cube for different size ratios $\alpha=1.2, 2.4$ and 5.4.}
	\label{fig:6}
\end{figure}

In Figure~\ref{fig:4}, we report the average cluster size for different size ratios $\alpha=\sigma_s/\sigma_c$ for the random parking algorithm based on selecting a random site for a sphere with diameter $\sigma_s$ on the surface of a cube with edge length $\sigma_c$, as described in the Methods section. The average cluster sizes fully resemble the ones reported in the main text for the experiments and the Monte Carlo simulations using  screened Coulomb  potentials for the interactions. The cluster size distributions, shown in Figure~\ref{fig:5}, also resemble  the ones extracted from Monte Carlo simulations and the experimental ones. This confirms the preference of the spheres to be located on the faces of the cube, as the probability to select a random position at one of the faces is higher than selecting a random site on one of the edges or corners of the cube. Therefore, the difference in cluster size distribution  found with the random adsorption parking algorithm  is due to the fact that the minimum distance condition between the surface of a cube and a sphere is enforced. Representative configurations for the random site parking algorithm  are reported in Figure~\ref{fig:6}.

\section*{Monte Carlo simulations with Yukawa interactions}

\begin{figure}[h!]
	\centering
	\includegraphics [width=0.6\textwidth]{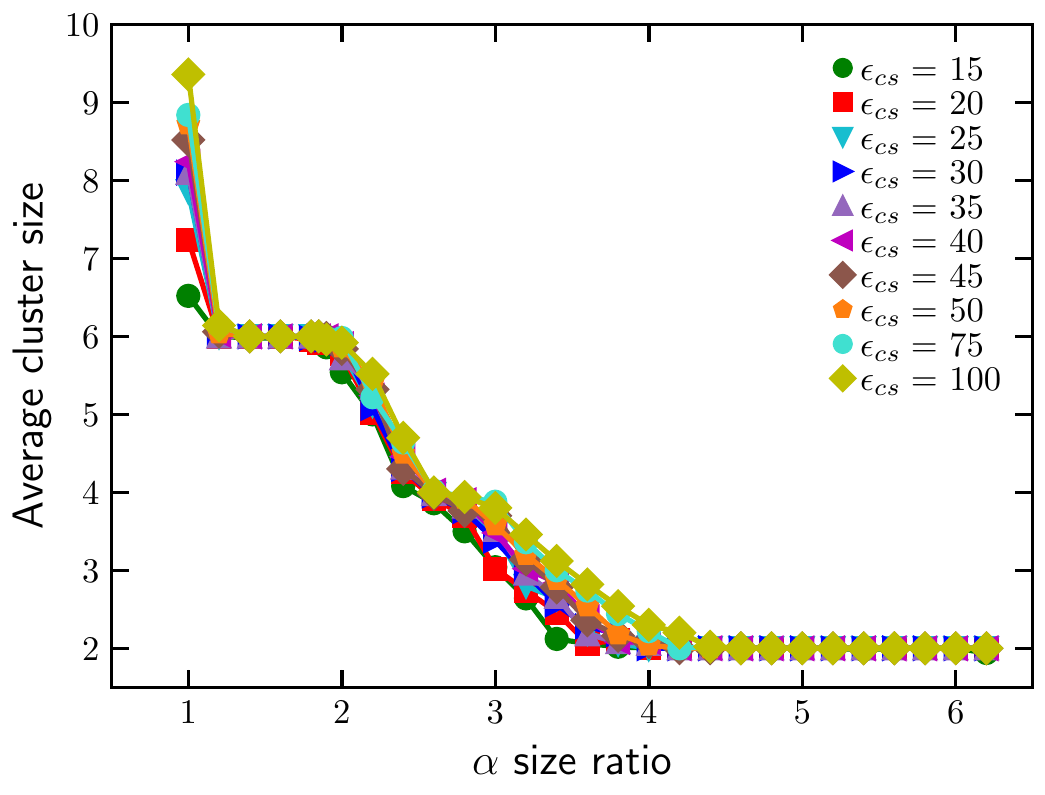} 
	\caption{Average number of particles attached to the central cube as a function of  size ratio $\alpha$ for varying values of the interaction strength $\epsilon_{cs}$.}
	\label{fig:7}
\end{figure}

In the main text, we show results as obtained from  Monte Carlo simulations using screened Coulomb interactions between the satellite spheres and the cubes  with interaction strengths $\epsilon_{cs}=15$ and $\epsilon_s=10$. In Figure~\ref{fig:7}, we show the average cluster sizes, i.e. average number of particles attached to the central cube, as a function of $\alpha$ for varying values of $\epsilon_{cs}$. We observe that the average cluster size has a very limited effect by changing the interaction strength $\epsilon_{cs}$. 

\clearpage
\newpage
\bibliography{Shelke_main}

\begin{thebibliography}{10}
\expandafter\ifx\csname url\endcsname\relax
  \def\url#1{\texttt{#1}}\fi
\expandafter\ifx\csname urlprefix\endcsname\relax\def\urlprefix{URL }\fi
\expandafter\ifx\csname href\endcsname\relax
  \def\href#1#2{#2} \def\path#1{#1}\fi

\bibitem{duguet2011design}
E.~Duguet, A.~D{\'e}sert, A.~Perro, S.~Ravaine, Design and elaboration of
  colloidal molecules: an overview, Chemical Society Reviews 40~(2) (2011)
  941--960.

\bibitem{poon2004colloids}
W.~Poon, Colloids as big atoms, Science 304~(5672) (2004) 830--831.

\bibitem{he2020colloidal}
M.~He, J.~P. Gales, {\'E}.~Ducrot, Z.~Gong, G.-R. Yi, S.~Sacanna, D.~J. Pine,
  Colloidal diamond, Nature 585~(7826) (2020) 524--529.

\bibitem{aryana2019superstructures}
K.~Aryana, J.~B. Stahley, N.~Parvez, K.~Kim, M.~B. Zanjani, Superstructures of
  multielement colloidal molecules: efficient pathways to construct
  reconfigurable photonic and phononic crystals, Advanced Theory and
  Simulations 2~(5) (2019) 1800198.

\bibitem{soto2014self}
R.~Soto, R.~Golestanian, Self-assembly of catalytically active colloidal
  molecules: tailoring activity through surface chemistry, Physical review
  letters 112~(6) (2014) 068301.

\bibitem{ni2017hybrid}
S.~Ni, E.~Marini, I.~Buttinoni, H.~Wolf, L.~Isa, Hybrid colloidal microswimmers
  through sequential capillary assembly, Soft Matter 13~(23) (2017) 4252--4259.

\bibitem{alvarez2021reconfigurable}
L.~Alvarez, M.~A. Fernandez-Rodriguez, A.~Alegria, S.~Arrese-Igor, K.~Zhao,
  M.~Kr{\"o}ger, L.~Isa, Reconfigurable artificial microswimmers with internal
  feedback, Nature Communications 12~(1) (2021) 1--9.

\bibitem{lowen2018active}
H.~L{\"o}wen, Active colloidal molecules, EPL (Europhysics Letters) 121~(5)
  (2018) 58001.

\bibitem{sun2015controllable}
Y.~Sun, M.~Chen, S.~Zhou, J.~Hu, L.~Wu, Controllable synthesis and surface
  wettability of flower-shaped silver nanocube-organosilica hybrid colloidal
  nanoparticles, ACS nano 9~(12) (2015) 12513--12520.

\bibitem{tian2022particle}
L.~Tian, Y.~Liu, D.~Wang, J.~Tan, Y.~Xie, B.~Li, Q.~Zhang, C.~Zhu, J.~Xu,
  Particle-click-particle: colloidal clusters from click seeded emulsion
  polymerization, Polym. Chem. 13 (2022) 1084--1089.

\bibitem{schamel2013chiral}
D.~Schamel, M.~Pfeifer, J.~G. Gibbs, B.~Miksch, A.~G. Mark, P.~Fischer, Chiral
  colloidal molecules and observation of the propeller effect, Journal of the
  American Chemical Society 135~(33) (2013) 12353--12359.

\bibitem{cho2005colloidal}
Y.-S. Cho, G.-R. Yi, S.-H. Kim, D.~J. Pine, S.-M. Yang, Colloidal clusters of
  microspheres from water-in-oil emulsions, Chemistry of materials 17~(20)
  (2005) 5006--5013.

\bibitem{ku2015soft}
K.~H. Ku, Y.~Kim, G.-R. Yi, Y.~S. Jung, B.~J. Kim, Soft patchy particles of
  block copolymers from interface-engineered emulsions, ACS nano 9~(11) (2015)
  11333--11341.

\bibitem{gong2017patchy}
Z.~Gong, T.~Hueckel, G.-R. Yi, S.~Sacanna, Patchy particles made by colloidal
  fusion, Nature 550~(7675) (2017) 234--238.

\bibitem{kurka2020self}
D.~W. Kurka, M.~Niehues, B.~J. Ravoo, Self-assembly of colloidal molecules
  based on host--guest chemistry and geometric constraints, Langmuir 36~(14)
  (2020) 3924--3931.

\bibitem{kraft2012surface}
D.~J. Kraft, R.~Ni, F.~Smallenburg, M.~Hermes, K.~Yoon, D.~A. Weitz, A.~van
  Blaaderen, J.~Groenewold, M.~Dijkstra, W.~K. Kegel, Surface roughness
  directed self-assembly of patchy particles into colloidal micelles,
  Proceedings of the National Academy of Sciences 109~(27) (2012) 10787--10792.

\bibitem{soto2002controlled}
C.~M. Soto, A.~Srinivasan, B.~R. Ratna, Controlled assembly of mesoscale
  structures using dna as molecular bridges, Journal of the American Chemical
  Society 124~(29) (2002) 8508--8509.

\bibitem{chakraborty2022self}
I.~Chakraborty, D.~J. Pearce, R.~W. Verweij, S.~C. Matysik, L.~Giomi, D.~J.
  Kraft, Self-assembly dynamics of reconfigurable colloidal molecules, ACS nano
  16~(2) (2022) 2471--2480.

\bibitem{chen2011supracolloidal}
Q.~Chen, J.~K. Whitmer, S.~Jiang, S.~C. Bae, E.~Luijten, S.~Granick,
  Supracolloidal reaction kinetics of janus spheres, Science 331~(6014) (2011)
  199--202.

\bibitem{sacanna2010lock}
S.~Sacanna, W.~T. Irvine, P.~M. Chaikin, D.~J. Pine, Lock and key colloids,
  Nature 464~(7288) (2010) 575--578.

\bibitem{maansson2019preparation}
L.~K. M{\aa}nsson, T.~De~Wild, F.~Peng, S.~H. Holm, J.~O. Tegenfeldt,
  P.~Schurtenberger, Preparation of colloidal molecules with
  temperature-tunable interactions from oppositely charged microgel spheres,
  Soft Matter 15~(42) (2019) 8512--8524.

\bibitem{demirors2015opposite}
A.~F. Demir\"ors, J.~C.~P. Stiefelhagen, T.~Vissers, F.~Smallenburg,
  M.~Dijkstra, A.~Imhof, A.~van Blaaderen, Long-ranged oppositely charged
  interactions for designing new types of colloidal clusters, Phys. Rev. X 5
  (2015) 021012.

\bibitem{mihut2017assembling}
A.~M. Mihut, B.~Stenqvist, M.~Lund, P.~Schurtenberger, J.~J. Crassous,
  Assembling oppositely charged lock and key responsive colloids: A mesoscale
  analog of adaptive chemistry, Science advances 3~(9) (2017) e1700321.

\bibitem{liu2021assembly}
Y.~Liu, J.~Wang, I.~Imaz, D.~Maspoch, Assembly of colloidal clusters driven by
  the polyhedral shape of metal--organic framework particles, Journal of the
  American Chemical Society 143~(33) (2021) 12943--12947.

\bibitem{bianchi2017limiting}
E.~Bianchi, B.~Capone, I.~Coluzza, L.~Rovigatti, P.~D. van Oostrum, Limiting
  the valence: advancements and new perspectives on patchy colloids, soft
  functionalized nanoparticles and biomolecules, Physical Chemistry Chemical
  Physics 19~(30) (2017) 19847--19868.

\bibitem{li2020colloidal}
W.~Li, H.~Palis, R.~Merindol, J.~Majimel, S.~Ravaine, E.~Duguet, Colloidal
  molecules and patchy particles: Complementary concepts, synthesis and
  self-assembly, Chemical Society Reviews 49~(6) (2020) 1955--1976.

\bibitem{ni2016programmable}
S.~Ni, J.~Leemann, I.~Buttinoni, L.~Isa, H.~Wolf, Programmable colloidal
  molecules from sequential capillarity-assisted particle assembly, Science
  advances 2~(4) (2016) e1501779.

\bibitem{wang2012colloids}
Y.~Wang, Y.~Wang, D.~R. Breed, V.~N. Manoharan, L.~Feng, A.~D. Hollingsworth,
  M.~Weck, D.~J. Pine, Colloids with valence and specific directional bonding,
  Nature 491~(7422) (2012) 51--55.

\bibitem{schade2013tetrahedral}
N.~B. Schade, M.~C. Holmes-Cerfon, E.~R. Chen, D.~Aronzon, J.~W. Collins, J.~A.
  Fan, F.~Capasso, V.~N. Manoharan, Tetrahedral colloidal clusters from random
  parking of bidisperse spheres, Phys. Rev. Lett. 110 (2013) 148303.

\bibitem{miracle_influence_2003}
D.~B. Miracle, W.~S. Sanders, O.~N. Senkov, The influence of efficient atomic
  packing on the constitution of metallic glasses, Philosophical Magazine
  83~(20) (2003) 2409--2428.

\bibitem{hu2020particle}
M.~Hu, C.-P. Hsu, L.~Isa, Particle surface roughness as a design tool for
  colloidal systems, Langmuir 36~(38) (2020) 11171--11182.

\bibitem{tagliazucchi_kinetically_2014}
M.~Tagliazucchi, F.~Zou, E.~A. Weiss, Kinetically {Controlled}
  {Self}-{Assembly} of {Latex}–{Microgel} {Core}–{Satellite} {Particles},
  The Journal of Physical Chemistry Letters 5~(16) (2014) 2775--2780.

\bibitem{manoharan2003dense}
V.~N. Manoharan, M.~T. Elsesser, D.~J. Pine, Dense packing and symmetry in
  small clusters of microspheres, Science 301~(5632) (2003) 483--487.

\bibitem{mage2019shape}
P.~L. Mage, A.~T. Csordas, T.~Brown, D.~Klinger, M.~Eisenstein, S.~Mitragotri,
  C.~Hawker, H.~T. Soh, Shape-based separation of synthetic microparticles,
  Nature materials 18~(1) (2019) 82--89.

\bibitem{hueckel2018mix}
T.~Hueckel, S.~Sacanna, Mix-and-melt colloidal engineering, ACS Nano 12~(4)
  (2018) 3533--3540.

\bibitem{wel2017surfactant}
C.~van~der Wel, N.~Bossert, Q.~J. Mank, M.~G.~T. Winter, D.~Heinrich, D.~J.
  Kraft, Surfactant-free colloidal particles with specific binding affinity,
  Langmuir 33~(38) (2017) 9803--9810.

\bibitem{sugimoto1992preparation}
T.~Sugimoto, K.~Sakata, Preparation of monodisperse pseudocubic $\alpha$-fe2o3
  particles from condensed ferric hydroxide gel, Journal of colloid and
  interface science 152~(2) (1992) 587--590.

\bibitem{dijkstra2002phase}
M.~Dijkstra, Phase behavior of hard spheres with a short-range yukawa
  attraction, Physical Review E 66~(2) (2002) 021402.

\bibitem{mansfield1996random}
M.~L. Mansfield, L.~Rakesh, D.~A. Tomalia, The random parking of spheres on
  spheres, The Journal of chemical physics 105~(8) (1996) 3245--3249.

\bibitem{tonti2021fast}
L.~Tonti, A.~Patti, Fast overlap detection between hard-core colloidal cuboids
  and spheres. the ocsi algorithm, Algorithms 14~(3) (2021) 72.

\bibitem{van_ravensteijn_colloids_2014}
B.~G.~P. van Ravensteijn, W.~K. Kegel, Colloids with {Continuously} {Tunable}
  {Surface} {Charge}, Langmuir 30~(35) (2014) 10590--10599.

\bibitem{ozaki1986reversible}
M.~Ozaki, H.~Suzuki, K.~Takahashi, E.~Matijevi{\'c}, Reversible ordered
  agglomeration of hematite particles due to weak magnetic interactions,
  Journal of colloid and interface science 113~(1) (1986) 76--80.

\bibitem{chen2011directed}
Q.~Chen, S.~C. Bae, S.~Granick, Directed self-assembly of a colloidal kagome
  lattice, Nature 469~(7330) (2011) 381--384.

\bibitem{glotzer2007anisotropy}
S.~C. Glotzer, M.~J. Solomon, Anisotropy of building blocks and their assembly
  into complex structures, Nature materials 6~(8) (2007) 557--562.

\bibitem{verweij2020flexibility}
R.~W. Verweij, P.~G. Moerman, N.~E. Ligthart, L.~P. Huijnen, J.~Groenewold,
  W.~K. Kegel, A.~van Blaaderen, D.~J. Kraft, Flexibility-induced effects in
  the brownian motion of colloidal trimers, Physical Review Research 2~(3)
  (2020) 033136.

\end{thebibliography}

\end{document}